\documentclass[lettersize,journal]{IEEEtran}
\newtheorem{Remark}{\it Remark}[section]

\newtheorem{Proposition}{\it Proposition}[section]
\newtheorem{Lemma}{\it Lemma}[section]

\makeatletter
\newcommand{\Rmnum}[1]{\expandafter\@slowromancap\romannumeral #1@}
\makeatother
\usepackage{booktabs} 
\usepackage{color}
\usepackage{graphicx} 
\usepackage{tabu}
\usepackage{pifont}
\usepackage{multicol}
\usepackage{setspace}
\usepackage{subfigure}
\usepackage{amsmath}
\usepackage{epstopdf}
\usepackage{indentfirst}
\usepackage{enumerate}
\usepackage{amssymb}
\usepackage{stfloats}
\usepackage{bm}
\usepackage{threeparttable}
\usepackage{CJK}
\usepackage[justification=centering]{caption}
\usepackage{makecell}
\usepackage{caption}
\usepackage{fancyhdr}
\usepackage{cases}
\usepackage{algorithm}
\usepackage{algpseudocode}
\usepackage{amsmath}
\usepackage{graphics}
\usepackage{subfig}
\usepackage{epsfig}
\usepackage{cite}

\begin{document}
\title{Joint Task and Data Oriented Semantic Communications: A Deep Separate Source-channel Coding Scheme}
\author{
	{Jianhao~Huang, Dongxu~Li, Chuan~Huang,  Xiaoqi~Qin, ~\IEEEmembership{Member,~IEEE}, and Wei Zhang, ~\IEEEmembership{Fellow,~IEEE} }
	\thanks{
	
	{Part of this work has been accepted by IEEE International Conference on Communications (ICC) 2023 \cite{huang_conf}. This work was supported in part by the Natural Science Foundation of China under Grant No. 62022070 and No. 62341112, in part by Key Area Research and Development Program of Guangdong Province under grant No. 2020B0101110003, in part by the key project of Shenzhen No. JCYJ20220818103006013, in part by the Shenzhen Outstanding Talents Training Fund 202002,  in part by the Guangdong Provincial Key Laboratory of Future Networks of Intelligence (Grant No. 2022B1212010001), in part by the Shenzhen Key Laboratory of Big Data and Artificial Intelligence (Grant No. ZDSYS201707251409055), and in part by Young Elite Scientists Sponsorship Program by China Association for Science and Technology 2021QNRC001. (Corresponding author: Chuan Huang).}
	
		{J. Huang, D. Li, and C. Huang are with the School of Science and Engineering and the Future Network of Intelligence Institute, the Chinese University of Hong Kong, Shenzhen, 518172 China. C. Huang is also with Peng Cheng Laboratory, Shenzhen, China, 518066. Emails: jianhaohuang1@link.cuhk.edu.cn, dongxuli@link.cuhk.edu.cn,  and huangchuan@cuhk.edu.cn. }
		
		{X. Qin is with the State Key Laboratory of Networking and Switching
Technology, Beijing University of Posts and Telecommunications, Beijing
100876, China. Email: xiaoqiqin@bupt.edu.cn.}

		{W. Zhang is with School of Electrical Engineering and Telecommunications, University of New South Wales, Sydney 4385, Australia. Email:  w.zhang@unsw.edu.au. }

	}
}
\maketitle

\IEEEpubid{\begin{minipage}{\textwidth}\ \\[30pt] \centering
		Copyright \copyright 2023 IEEE. Personal use of this material is permitted. 
		However, permission to use this material for any other purposes must \\ be obtained 
		from the IEEE by sending an email to pubs-permissions@ieee.org.
\end{minipage}}

\begin{abstract}
Semantic communications are expected to accomplish various semantic tasks with relatively less spectrum resource by exploiting the semantic feature of source data. To simultaneously serve both the data transmission and semantic tasks, joint data compression and semantic analysis has become pivotal issue in semantic communications. This paper proposes a deep separate source-channel coding (DSSCC) framework for the joint task and data oriented semantic communications (JTD-SC) and utilizes the variational autoencoder approach to solve the rate-distortion problem with semantic distortion.  First, by analyzing the Bayesian model of the DSSCC framework, we derive a novel rate-distortion optimization problem via the Bayesian inference approach for general data distributions and semantic tasks. Next, for a typical application of joint image transmission and classification, we combine the variational autoencoder approach with a forward adaption scheme to effectively extract image features and adaptively learn the density information of the obtained features. Finally,  an iterative training algorithm is proposed to tackle the overfitting issue of deep learning models.  Simulation results reveal that the proposed scheme achieves better coding gain as well as data recovery and classification performance in most scenarios, compared to the classical compression schemes and the emerging deep joint source-channel schemes.

\end{abstract}


\section{Introduction}
With the advent of the fifth-generation (5G)  communication era, the explosive growth of multimedia applications, e.g., extended reality, autonomous driving, and intelligent surveillance, poses tremendous challenges on the utilization of limited spectrum resources, which promotes the evolution from bit communications \cite{Tse2005} to semantic communications \cite{zhang2022toward,kim2019new}. Empowered by the innovations of artificial intelligence (AI), semantic communications start a new paradigm to extract, encode, and transmit the semantic information of source data, e.g., image feature, object labels, and attributes \cite{zhang2022toward}, rather than  to simply transmit the data itself.  By this mean, various machine tasks, e.g., pedestrian monitoring, defect detection, and security surveillance \cite{liu2022task, 9398576}, can be efficiently accomplished with  relatively less spectrum resources. 
However, in some internet of things (IoT) scenarios, e.g. real-time surveillance \cite{IoT1}, semantic communications need to simultaneously serve both the data transmissions and certain semantic tasks, which poses new challenges on the joint data compression and semantic analysis.

For the joint task and data oriented semantic communications (JTD-SC), conventional communication system follow the ``reconstruct-and-then-analyze'' paradigm, where source data is  first  compressed into bit streams and then transmitted to the receiver for reconstruction, and finally the reconstructed data is used to accomplish the semantic tasks. The reconstruction distortion is usually measured by the mean square error (MSE) for  image and video data \cite{wallace1992jpeg,Balle2017,stuhlmuller2000analysis}, bilingual evaluation understudy (BLEU) for  text data \cite{jiang2022deep}, etc. With this paradigm, a large volume of data compression methods \cite{wallace1992jpeg,christopoulos2000jpeg2000, stuhlmuller2000analysis, jiang2022deep,sullivan2012overview,Balle2017,Balle2018,9242247,theis2017lossy}
have been proposed and aim to achieve the tradeoffs between the coding rate and the reconstruction  distortion. For example, the widely-used image lossy compression schemes Joint Photographic Experts Group (JPEG) \cite{wallace1992jpeg}  and JPEG2000 \cite{christopoulos2000jpeg2000} utilize the Discrete Cosine Transform (DCT) and Discrete Wavelet Transform (DWT), respectively, to transform the images into the frequency domain, followed by quantization, entropy coding, and channel coding. By utilizing intra-frame coding in video coding standard, Better Portable Graphics (BPG) \cite{sullivan2012overview} has been shown to achieve higher compression efficiency than JPEG and JPEG2000. However, the above hand-craft schemes utilize either linear  or fixed transform functions, which cannot capture the distribution information of source data. 

Recently, deep neural network (DNN)  has become more and more compelling in the field of data compression, due to its low computational complexity and high capability of approximating nonlinear functions \cite{8054694,Balle2017,Balle2018,9242247,theis2017lossy}.  It turns out that combining with the stochastic optimization tools, e.g., stochastic gradient descent (SGD), and the rate-distortion loss function, the DNN-based scheme can effectively eliminate the redundancy of source data and achieve higher compression efficiency compared with the hand-craft schemes \cite{wallace1992jpeg,christopoulos2000jpeg2000}.   The design principles of the  DNN-based compression schemes can be divided into two categories: deep separate source-channel coding (DSSCC)\cite{Balle2017,Balle2018,9242247,theis2017lossy} and deep joint source-channel coding (DJSCC) \cite{kurka2020deepjscc,dai2022nonlinear,bourtsoulatze2019deep}. For the former one,  the traditional lossy compression scheme, e.g., JPEG, is replaced with the DNN architecture, followed by quantization and entropy coding. Following this idea, the authors in \cite{Balle2017} proposed a variational autoencoder method to jointly optimize the  encoder and decoder  by approximating the quantization error as uniform noise. To achieve higher coding gain, recent works applied hyperpriors into the auoencoder method to model the probability density function (PDF) of the semantic features more accurately \cite{Balle2018,9242247}. In addition, generative adversarial network (GAN) \cite{creswell2018generative} was utilized to generate the data more naturally close to the original ones by training the adversarial network at the receiver.  For the DJSCC scheme, the source and channel codings are integrated by utilizing the DNN architecture  and the data is directly transformed into continuous-valued symbols for transmissions. The landmark works \cite{kurka2020deepjscc,dai2022nonlinear,bourtsoulatze2019deep} on the DJSCC schemes have shown the higher compression efficiency than the classical separation-based JPEG/JPEG2000/BPG compression schemes combined with ideal channel capacity-achieving code over some image datasets. However, the lack of quantization and constellation diagrams in the DJSCC scheme may make it less compatible with modern communication hardware and standardized protocols. In addition, the DJSCC scheme relies on the off-line training and suffers from severe performance degradation when channel conditions change.
After recovering the source data, lots of the DNN-based schemes can be utilized to extract the semantic information for different tasks, e.g., action recognition \cite{wang2013action}, object recognition \cite{krizhevsky2017imagenet}, image captioning \cite{hossain2019comprehensive}, etc.
However, neither the above DSSCC nor the DJSCC data compression method considers  the semantic information of data, and thus the recovered data under high compression ratio will seriously degrade the performance of semantic tasks \cite{roy2018effects}.

To improve the performance of semantic tasks, an alternative way is to extract and compress the semantic information and then transmit it  over independent channels \cite{shao2021learning,wang2013action,hossain2019comprehensive,krizhevsky2017imagenet,tu2021semantic}, which may introduce extra communication cost.  Furthermore, extracting  semantic information is usually resource consuming, which will cause severe computational overhead for the front-end devices.  Hence, it is essential to conduct the joint design of data compression and semantic analysis. The authors in \cite{liu2022task}  proposed a novel DJSCC based scheme to reconstruct data with task semantics by using the information bottleneck method.  For the DSSCC scheme, multi-task learning techniques are usually employed to address joint data compression and semantic recovery, where the total loss function is based on a weighted summation of the individual task-specific loss functions. For example, the authors in \cite{luo2018deepsic} proposed an autoencoder scheme with the deep semantic image compression (DeepSIC) model to compress and reconstruct both the image data and its label information, where the rate-distortion function for training the DNNs was extended by adding the semantic distortion. Based on the DeepSIC model, the authors in  \cite{patwa2020semantic} proposed  a modified DNN architecture to improve the image classification performance. The authors in \cite{kawawa2022recognition} combined Gaussian mixture model with EfficientNet-B0 \cite{tan2019efficientnet} for joint image compression and recognition. Although the aforementioned DSSCC schemes investigate diverse DNN architectures to enhance the joint data compression and semantic recovery performance, they do not consider the Bayesian logic of the overall system. Moreover, the intricate relationship between the variational autoencoder and the rate-distortion theory with semantic distortion remains unexplored.

 The purpose of this paper  is to propose an efficient DSSCC-based approach for the JTD-SC problem, by  exploring the rate-distortion theory with semantic distortion.  In particular,  we consider the JTD-SC problem over the point-to-point channel, where the transmitter aims to extract the low-dimensional features of the source data and sends it to the receiver for recovering both the data and its semantic information. The semantic information is regarded as the latent random variable of the source data and is estimated from the recovered data by applying the optimal maximum a posterior probability (MAP) scheme.  Different from the conventional data compression schemes \cite{wallace1992jpeg,christopoulos2000jpeg2000, stuhlmuller2000analysis, jiang2022deep,sullivan2012overview,Balle2017}, the key idea of the proposed variational autoencoder scheme is to jointly optimize the feature extraction function and data recovery function to achieve the trade-offs between the coding rate and the distortions of both the data and semantic information. 
 By exploiting the distributions of the source data and semantic information,  the performance of both the data transmissions and its semantic information recovery can be improved.
 The main contributions of this paper are summarized as follows:

\begin{itemize}
	\item First, we propose a DSSCC-based framework  for the JTD-SC  with the general source data and semantic tasks.  By exploiting the Bayesian model of the DSSCC-based framework, we derive a novel rate-distortion optimization problem for the JTD-SC problem via the variational autoencoder approach, which is shown to achieve the trade-offs between the coding rate and the distortions of both the data and semantic information.  Compared with the extended rate-distortion problem \cite{luo2018deepsic} only for image data, the proposed rate-distortion optimization problem utilizes the conditional entropy to express the distortions of general data and semantic information.
		
\item	Then, we take the image transmission and classification task as example to implement the proposed design framework by solving the proposed rate-distortion optimization problem.
 Particularly, by integrating the \emph{forward adaption} (FA) method \cite{dai2022nonlinear} and the variational autoencoder approach, we propose an FA-based autoencoder scheme to improve the performance of  both the image transmission and semantic information recovery. The key idea of the FA scheme is to learn the parametric density model as side information to approximate the distribution of  image feature, which usually has no closed-form expression. 
	
	\item Finally, we propose a training algorithm to iteratively train the DNNs for data recovery and semantic tasks, in which the training batch sizes for the semantic tasks can be flexibly changed to tackle the overfitting problem.  Simulation results reveal that the proposed scheme with either capacity achieving channel code \cite{markov,gallager1968information} or low density parity check (LDPC) code \cite{gallager1968information} can achieve better image recovery and classification performances in most scenarios,  compared to the conventional data compression methods, e.g., BPG, JPEG, and DJSCC schemes.
	\end{itemize}

The reminder of this paper is organized as follows. Section II introduces the DSSCC-based framework for the JTD-SC problem over the point-to-point channel. Section III
derives the rate-distortion optimization problem via the variational autoencoder scheme.
Section IV proposes  an FA-based autoencoder scheme for the image transmission and classification, and presents the implementation details.
Section V shows the simulation results. Finally, Section VI summarizes this paper. 

Notations:  We use lowercase and uppercase letters, e.g., $x$ and $X$, to denote  scalars, and use boldface lowercase letters, e.g., $\bm{x}$, to denote vectors. $\mathbb{Z}$, $\mathbb{C}$,  and $\mathbb{R}$ denote the sets of all integer, complex, and  real values, respectively.
$||\bm{x}||$  and $||\bm{x}||_1$ denote the $2$-norm and $1$-norm of vector $\bm{x}$, respectively.
$\bm{x}^T$ and $\bm{x}^H$ denote the  transpose and conjugate transpose of vector $\bm{x}$, respectively.  $p_{\bm{x}}(\bm{x})$ denotes the PDF of the continuous random variable $\bm{x}$.  $P_{\bm{y}}(\bm{y})$ denotes the probability mass function (PMF) of the discrete random variable $\bm{y}$. 
$\mathbb{E}_{\bm{y}\sim p_{\bm{y}}}(\cdot)$  denotes the expectation for  random variable ${\bm{y}}$ with the PDF being $p_{\bm{y}}(\bm{y})$.   $\text{log}(\cdot)$ is the logarithm function with base $2$. $[L]$ denotes the set  of the positive integers not bigger than  $L$.

%It still remains an unsolved problem how to measure the performance metric of the image recovery. Most of the paper used the mean square error or SSIM to measure the performance of the image compression and recovery, and these metrics are based on the human perceptual sensitivity. However, in the 5G communications, images are also used for certain tasks, such as object classification and recognition, and the previous metrics may not be suitable for these tasks. In these cases, the tradeoff between the image recovery and semantic information transmission is still mystery territory needed to be explored. 
\section{System Model}
 \begin{figure*}[t]
\normalsize
\setlength{\abovecaptionskip}{+0.3cm}
\setlength{\belowcaptionskip}{-0.1cm}
\centering
\includegraphics[width=5.0in]{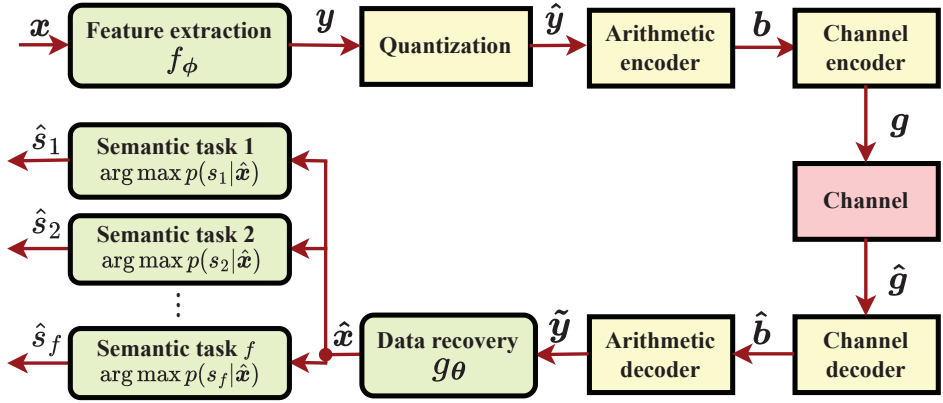}
%\caption{fig2}
\caption{Framework of the considered DSSCC scheme.}
\label{figure:one}
\end{figure*}

Consider a JTD-SC problem for the point-to-point channel where the transmitter aims to compress the $M$-dimensional data vector $\bm{x}$ and sends it to the  receiver for recovery and processing certain semantic tasks. To save the transmission bandwidth,  data $\bm{x}$ needs to be lossily compressed while maintaining sufficient semantic fidelity for various semantic tasks. Here, processing semantic tasks is referred as recovering the corresponding semantic information $\bm{s}$ of data $\bm{x}$. For example, image classification task refers to recovering image's label information $\bm{s} \in \mathbb{S}$, where $\mathbb{S}$ is the set containing all the possible image labels. 

With the above setup, a transmission and reception framework based on the DSSCC scheme is proposed, as shown in Fig. \ref{figure:one}. At the transmitter, the compression and transmission processes are excuted via a pipeline which breaks down into four modules: feature extraction, quantization, arithmetic coding \cite{witten1987arithmetic}, and channel coding.
First, the input data $\bm{x} \in \mathbb{R}^{M}$, whose distribution is denoted as $p_{\bm{x}}(\bm{x})$, is mapped into a $K$-dimensional vector $\bm{y}$ by the feature extraction function $f_{\bm{\phi}}$ with parameter $\bm{\phi}$. For lossy compression, $K$ is much smaller than $M$.  Then,  continuous vector $\bm{y}$ is quantized as  discrete vector $\hat{\bm{y}}\in \mathbb{R}^{K}$, i.e., $\hat{\bm{y}}=\left\lceil \bm{y} \right\rfloor$, where $\left\lceil \cdot \right\rfloor$ denotes the uniform scalar quantization operation \cite{9242247}. Next,  lossless entropy coding, i.e.,  arithmetic coding, is employed to encode the quantized vector  $\hat{\bm{y}}$ into  bit stream $\bm{b}$ according to its PMF $P_{\hat{\bm{y}}}(\hat{\bm{y}})$. The expected coding rate of vector $\hat{\bm{y}}$ is calculated as $\mathbb{E}_{\bm{x}}\{-\log P_{\hat{\bm{y}}}(\hat{\bm{y}}) \}$. Finally,  bit sequence $\bm{b}$ is encoded by the channel encoder into  complex symbol vector $\bm{g}\in \mathbb{C}^{L}$ with  length $L$ for transmissions.  Here, the channel bandwidth ratio \cite{bourtsoulatze2019deep} is defined as  $L/M$ to measure the averaged bandwidth cost for transmitting one element of data $\bm{x}$. 

The input-and-output relationship of the point-to-point channel is expressed as 
\begin{align}\label{signal}
	\hat{g}_j=hg_j+n_j,\  j \in[L],
\end{align}
where ${\bm{g}}=[{g}_1,{g}_2,\cdots,{g}_L]^T$, $\hat{\bm{g}}=[\hat{g}_1,\hat{g}_2,\cdots,\hat{g}_L]^T$ denotes the received symbols, and $\bm{n}=[n_1,n_2,\cdots,n_L]^T$ is the independent and identically distributed (i.i.d.) 
circularly symmetric complex Gaussian (CSCG) noise with  mean zero and variance $ \delta_n^2\bm{I}$. $h\in \mathbb{C}$ is the constant channel coefficient over $L$ channel uses and is known to the receiver. Then, the average signal-to-noise ratio (SNR) of the considered channel is defined as $\frac{\mathbb{E}(\bm{g}^H\bm{g})}{L\delta_n^2}$.

At the receiver side, the received symbols $\hat{\bm{g}}$ are decoded and dequantized as the feature vector $\tilde{\bm{y}}$. Then, vector $\tilde{\bm{y}}$ is fed into the  recovery function $g_{\bm{\theta}}$ with parameter $\bm{\theta}$ to recover the transmitted data $\bm{x}$. Finally, the recovered data $\hat{\bm{x}}$ is used to recover the semantic information $(\hat{\bm{s}}_1,\hat{\bm{s}}_2,\cdots,\hat{\bm{s}}_f)$ with $f$ being a positive integer. The semantic recovery follows the optimal MAP scheme \cite{markov,Tse2005}, i.e.,
\begin{align} \label{seman-rec}
	\hat{\bm{s}}_i=\arg \max_{\bm{s}_i \in \mathbb{S}_i} p_{\bm{s}_i|\hat{\bm{x}}}(\bm{s}_i|\hat{\bm{x}}), \ i \in [f].
\end{align}
where $\mathbb{S}_i$ is the set containing all the possible values of the semantic information $\bm{s}_i$. 
%In this paper, we take the widely-used image classification task \cite{krizhevsky2017imagenet} as an example, and regard the labels of images, e.g., boat, human, animals, and etc, as the semantic information that needs to be recovered. 

For the above system model, our goal is to jointly design the feature extraction function $f_{\bm{\phi}}$ and the  recovery function $g_{\bm{\theta}}$ to minimize the weighted sum of  the coding rate and the distortions of data and semantic information, which is formulated as the following problem
\begin{align} \label{rate-dis}
	(\bm{\phi}^*,\bm{\theta}^*)=\arg \min_{\{\bm{\phi},\bm{\theta}\}}  & \mathbb{E}_{\bm{x}}\{-\log P_{\hat{\bm{y}}}(\hat{\bm{y}}) \}+{\lambda}_0 d(\bm{x},\hat{\bm{x}})\nonumber \\
	&+\sum_{i=1}^{f}{\lambda}_i d_i(\bm{s}_i,\hat{\bm{s}}_i),
\end{align}
where $\lambda_0 \geq 0$, $\lambda_i \geq 0, i\in[f]$,  $ \mathbb{E}_{\bm{x}}\{-\log P_{\hat{\bm{y}}}(\hat{\bm{y}}) \}$ denotes the expected coding rate of feature $\hat{\bm{y}}$, $d(\bm{x},\hat{\bm{x}})$ denotes the distortion  between the original data $\bm{x}$ and the recovered data $\hat{\bm{x}}$, and $d_i(\bm{s}_i,\hat{\bm{s}}_i)$ denotes the distortion  between the original semantic information $\bm{s}_i$ and the recovered semantic information $\hat{\bm{s}}_i$. 

However, there are several challenges making the problem given in \eqref{rate-dis} difficult to be solved by the conventional statistical or optimization methods \cite{markov,boyd2004}: First,  PMF  $P_{\hat{\bm{y}}}(\hat{\bm{y}})$ in \eqref{rate-dis} is a discrete function, which makes it impossible to optimize the parameters $\{\bm{\phi},\bm{\theta}\}$ by applying the efficient gradient descent method \cite{boyd2004}; Second, the  distribution of  source data $\bm{x}$ is usually intractable, and thus the Bayesian estimation methods, e.g., MAP and minimum MSE (MMSE) \cite{markov}, cannot be directly applied to recover the data and the semantic information; Third, the posterior PDF $p_{\bm{s}_i|\hat{\bm{x}}}(\bm{s}_i|\hat{\bm{x}})$ is hard to be obtained due to the coupled parameters $\{\bm{\phi},\bm{\theta}\}$, and thus the distortion $d_{i}(\bm{s}_i,\hat{\bm{s}}_i)$ usually has no closed-form expression.

\section{Proposed Variational Autoencoder Approach}

This section proposes a variational autoencoder approach for the JTD-SC problem given in \eqref{rate-dis}, which only requires access to a sufficient number of data samples and not prior statistical knowledge of the source data. To tackle the challenges in problem \eqref{rate-dis}, the variational autoencoder approach first approximates  PMF $P_{\hat{\bm{y}}}(\hat{\bm{y}})$ by a continuous PDF function  and then utilizes the deep learning methods to jointly learn the parameters  $\{\bm{\phi},\bm{\theta}\}$. 
To facilitate the analysis, we first present the Bayesian model of the aforementioned communication system in Fig. \ref{figure:one}. Based on the obtained Bayesian model, we then derive a new formulation of the rate-distortion optimization problem in \eqref{rate-dis} by utilizing the variational inference method. Finally, we show how to compute the newly derived  optimization problem for general data distributions.

\subsection{Bayesian model}

\begin{figure}[htbp]
\centering

\subfigure[Inference model]{
\begin{minipage}[t]{1\linewidth}
	\centering
	\includegraphics[width=2.3in]{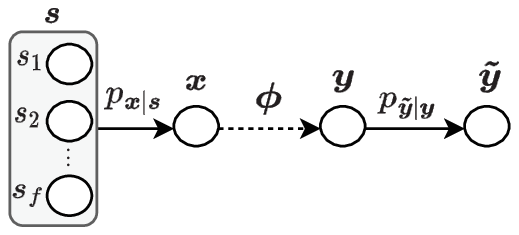}
%\caption{fig1}
\end{minipage}%
}%

\subfigure[Generative model]{
\begin{minipage}[t]{1\linewidth}
	\centering
	\includegraphics[width=2.3in]{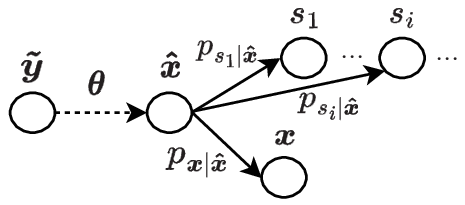}
%\caption{fig2}
\end{minipage}%
}%
\centering
\caption{Bayesian model of the considered semantic communication system. White circle represents random variable, solid line arrow represents the conditional dependence between random variables, and dotted line arrow represents the deterministic mapping between random variables with specific parameters.}
\label{Appendix:2}
\end{figure}

With the variational autoencoder approach \cite{Balle2017}, the compression and recovery processes in the considered  semantic communication system can be modeled as the inference  and  generative models, respectively, as shown in Fig. \ref{Appendix:2}. Following the Bayesian principle \cite{markov},  
the inference  and  generative models enable efficient ways to calculate the  conditional PDFs  $p_{\bm{\tilde{y}}|\bm{{x}}}(\bm{\tilde{y}}|\bm{{x}};\bm{\phi})$ and 
$p_{\bm{\tilde{y}}|\bm{s}_i}(\bm{\tilde{y}}|\bm{s}_i;\bm{\phi})$, and the posterior PDFs  $p_{\bm{{x}}|\bm{\tilde{y}}}(\bm{{x}}|\bm{\tilde{y}};\bm{\theta})$ and 
$p_{\bm{s}_i|\bm{\tilde{y}}}(\bm{s}_i|\bm{\tilde{y}};\bm{\theta})$, respectively, which are used for the design of parameters $\bm{\phi}$ and $\bm{\theta}$.

In Fig. \ref{Appendix:2}(a),  the inference model represents the Markov chain from  source data $\bm{x}$ to  feature $\tilde{\bm{y}}$. The semantic information $\bm{s}=[\bm{s}_1,\bm{s}_2,\cdots,\bm{s}_f]^T$ is regarded as the latent random variables of data $\bm{x}$, whose conditional PDF is denoted as $p_{\bm{x}|\bm{s}}(\bm{x}|\bm{s})$.  Function $f_{\bm{\phi}}$ with parameter $\bm{\phi}$
represents the inference mapping from vector $\bm{x}$
to vector ${\bm{y}}$. The conditional probability from vector $\bm{y}$ to vector $\tilde{\bm{y}}$ is denoted as $p_{\bm{\tilde{y}}|\bm{y}}(\tilde{\bm{y}}|\bm{y})$, which captures the distribution of errors caused by the quantization and signal transmissions in \eqref{signal}. For the sake of argument, we consider an ideal case that the transmission over the point-to-point channel is error-free\footnote{Error-free transmission means that  bit error probability of transmitting bit stream $\bm{b}$ over the channel is zero, i.e., $\hat{\bm{b}}=\bm{b}$, which implies $\hat{\bm{y}}=\tilde{\bm{y}}$. Theoretically, when the transmission rate is less than the channel capacity, i.e., $\frac{-\log P_{\hat{\bm{y}}}(\hat{\bm{y}})}{L}<\log (1+h^Hh\rho)$ with $\rho$ being the transmit SNR, it is proved that there always exists a channel code to make the bit error probability exponentially vanish with respect to the number of channel uses $L$ \cite{gallager1968information}. Notably, in practical scenarios, the transmission with certain bit error probability does result in performance degradation, which will be discussed in the simulation part.}\cite{Balle2017,Balle2018,9242247,theis2017lossy}, i.e., $\hat{\bm{b}}=\bm{b}$ and $\hat{\bm{y}}=\tilde{\bm{y}}$, and then signal $\bm{\tilde{y}}$ only suffers from the quantization error.  The quantization errors of the feature elements $\{\tilde{y}_i\}$ are approximated by the i.i.d. uniform distribution \cite{Balle2017}, i.e., 
\begin{align} \label{uniform}
	p_{\bm{\tilde{y}}|\bm{y}}(\bm{\tilde{y}}|\bm{y})=\prod \limits_{i=1}^{K}\mathcal{U}(\tilde{y}_i;y_i,1),
\end{align}
where $\bm{\tilde{y}}=[\tilde{y}_1,\tilde{y}_2,\cdots,\tilde{y}_K]^T$, $\bm{y}=[y_1,y_2,\cdots,y_K]^T$, and $\mathcal{U}(\tilde{y}_i;y_i,1)$ denotes the PDF of uniformly distributed random variable with the unit interval centered on $y_i$. Then, 
PDF $p_{\tilde{y}_i}(\tilde{y}_i), i \in [K],$ is calculated as the convolution of PDF $p_{y_i}(y_i)$ and  uniform distribution in \eqref{uniform}, i.e., 
\begin{align} \label{hyper:1}
	p_{\tilde{y}_i}(\tilde{y}_i)= p_{y_i}(y_i)*\mathcal{U}(\tilde{y}_i;y_i,1)=\int_{\tilde{y}_i-0.5}^{\tilde{y}_i+0.5} p_{y_i}(y_i) dy_i,
\end{align}
where $ i \in [K]$, $\bm{y}=f_{\bm{\phi}}(\bm{x})$, and $f(x)*h(x)$ denotes the convolution between functions $f(x)$ and $h(x)$. Given the inference model in Fig. \ref{Appendix:2}(a) and \eqref{uniform}, 
the conditional PDFs $p_{\bm{\tilde{y}}|\bm{{x}}}(\bm{\tilde{y}}|\bm{{x}};\bm{\phi})$ and 
$p_{\bm{\tilde{y}}|\bm{s}_i}(\bm{\tilde{y}}|\bm{s}_i;\bm{\phi})$ are calculated as 
\begin{align}
\label{variational1}	p_{\bm{\tilde{y}}|\bm{{x}}}(\bm{\tilde{y}}|\bm{{x}};\bm{\phi})&=\prod \limits_{i=1}^{K}\mathcal{U}(\tilde{y}_i;y_i,1),\\
\label{variational2}	p_{\bm{\tilde{y}}|\bm{s}_i}(\bm{\tilde{y}}|\bm{s}_i;\bm{\phi})&=\int_{\bm{x}} \prod \limits_{j=1}^{K}\mathcal{U}(\tilde{y}_j;y_j,1)p_{\bm{x}|\bm{s}_i}(\bm{x}|\bm{s}_i) d \bm{x},
\end{align}
with $\bm{y}=f_{\bm{\phi}}(\bm{x})$.

 Fig. \ref{Appendix:2}(b) represents the generative model that describes the recovery process at the receiver.
The recovery function $g_{\bm{\theta}}$ with parameter $\bm{\theta}$ represents the generative mapping from vector $\tilde{\bm{y}}$ to the recovered data $\bm{\hat{x}}$.  According to the generative model in Fig. \ref{Appendix:2}(b), the posterior PDFs $p_{\bm{{x}}|\bm{\tilde{y}}}(\bm{{x}}|\bm{\tilde{y}};\bm{\theta})$ and 
$p_{\bm{s}_i|\bm{\tilde{y}}}(\bm{s}_i|\bm{\tilde{y}};\bm{\theta})$  can also be formulated as 
\begin{align}
\label{equ:pos1}	p_{\bm{{x}}|\bm{\tilde{y}}}(\bm{{x}}|\bm{\tilde{y}};\bm{\theta})&=p_{\bm{x}|\bm{\hat{x}}}(\bm{{x}}|\bm{\hat{x}}),\\
\label{equ:pos2}	p_{\bm{s}_i|\bm{\tilde{y}}}(\bm{s}_i|\bm{\tilde{y}};\bm{\theta})&=p_{\bm{s}_i|\bm{\hat{x}}}(\bm{s}_i|\bm{\hat{x}}),\ \text{with}\  \bm{\hat{x}}=g_{\bm{\theta}}(\bm{\tilde{y}}).
\end{align}

According to the  Bayesian inference theory \cite{kingma2013auto}, it is expected that the design of parameters $\bm{\phi}$ and $\bm{\theta}$ should follow the Bayesian principle, i.e., 
\begin{align}
	p_{\bm{x}|\bm{\tilde{y}}}(\bm{x}|\bm{\tilde{y}};\bm{\theta})&=p_{\tilde{\bm{y}}|\bm{x}}(\tilde{\bm{y}}|\bm{x};\bm{\phi})p_{\bm{x}}(\bm{x})/p_{\bm{\tilde{y}}}(\bm{\tilde{y}}),\\
	p_{\bm{s}_i|\bm{\tilde{y}}}(\bm{s}_i|\bm{\tilde{y}};\bm{\theta})&=p_{\tilde{\bm{y}}|\bm{s}_i}(\tilde{\bm{y}}|\bm{s}_i;\bm{\phi})p_{\bm{s}_i}(\bm{s}_i)/p_{\bm{\tilde{y}}}(\bm{\tilde{y}}), i\in[f],
\end{align}which are generally intractable due to the unknown prior PDFs $p_{\bm{x}}(\bm{x})$ and $p_{\bm{s}}(\bm{s})$. To tackle this issue, the variational autoencoder approach \cite{Balle2017} aims to parameterize  functions $f_{\bm{\phi}}$ and $g_{\bm{\theta}}$ as DNNs and then approximates PDFs  $p_{\bm{\tilde{y}}|\bm{{x}}}(\bm{\tilde{y}}|\bm{{x}};\bm{\phi})$ and $p_{\bm{\tilde{y}}|\bm{s}_i}(\bm{\tilde{y}}|\bm{s}_i;\bm{\phi})$ given in  \eqref{variational1} and \eqref{variational2} to PDFs $p_{\bm{\tilde{y}}|\bm{{x}}}(\bm{\tilde{y}}|\bm{{x}};\bm{\theta})$ and 
$p_{\bm{\tilde{y}}|\bm{s}_i}(\bm{\tilde{y}}|\bm{s}_i;\bm{\theta})$ from \eqref{equ:pos1} and \eqref{equ:pos2}.

%Given the posterior PDFs $p_{\bm{{x}}|\bm{\tilde{y}}}(\bm{{x}}|\bm{\tilde{y}};\bm{\theta})$ and 
%$p_{\bm{s}_i|\bm{\tilde{y}}}(\bm{s}_i|\bm{\tilde{y}};\bm{\theta})$, we try to compute the conditional PDFs $p_{\bm{\tilde{y}}|\bm{{x}}}(\bm{\tilde{y}}|\bm{{x}};\bm{\theta})$ and 
%$p_{\bm{\tilde{y}}|\bm{s}_i}(\bm{\tilde{y}}|\bm{s}_i;\bm{\theta})$ with respect to (w.r.t.) the parameter $\bm{\theta}$, which are generally intractable due to the unknown prior PDFs $p_{\bm{x}}(\bm{x})$ and $p_{\bm{s}}(\bm{s})$. To this end
%It is worth noticing that the inference model and the generative model represent the opposed Markov chains. Conventionally speaking, the parameters $\bm{\theta}$ and $\bm{\phi}$ are highly correlated and  can be converted into each other when the prior distribution $p_{\bm{x}}(\bm{x})$ is available. For example, when the parameter $\bm{\phi}$ is fixed, the recovery process for data $\bm{x}$  should follow the optimal MAP principle by computing $p_{\bm{x}|\bm{\tilde{y}}}(\bm{x}|\bm{\tilde{y}};\bm{\theta})=p_{\bm{x}|\bm{\tilde{y}}}(\bm{x}|\bm{\tilde{y}};\bm{\phi})=p_{\tilde{\bm{y}}|\bm{x}}(\tilde{\bm{y}}|\bm{x};\bm{\phi})p_{\bm{x}}(\bm{x})/p_{\bm{\tilde{y}}}(\bm{\tilde{y}})$. However, the prior distribution $p_{\bm{x}}(\bm{x})$ is usually intractable in the practice situations, e.g., image and text, which makes it impossible to find the direct mapping between the parameters $\bm{\theta}$ and $\bm{\phi}$. 

\subsection{Proposed  approach}

\newcounter{TempEqCnt1}
\newcounter{TempEqCnt}
\setcounter{TempEqCnt1}{\value{equation}}
\setcounter{TempEqCnt}{\value{equation}}
	\begin{figure*}[b]
		  \hrulefill
		\vspace{-5pt}
		\setcounter{equation}{12}
		\begin{align} \label{equ:upper}
		\mathcal{L}_{\bm{\phi},\bm{\theta}}& \leq  (1+\lambda)\underbrace{\mathbb{E}_{\bm{x}} \mathbb{E}_{\tilde{\bm{y}}\sim q_{\tilde{\bm{y}}|\bm{x}}}\left\{-\log p_{\tilde{\bm{y}}}(\tilde{\bm{y}})\right\}}_{\mathcal{R}}+\underbrace{\mathbb{E}_{\bm{x}} \mathbb{E}_{\tilde{\bm{y}}\sim q_{\tilde{\bm{y}}|\bm{x}}} \left\{-\text{log} \ p_{\bm{x}|\bm{\hat{x}}}(\bm{x}|\bm{\hat{x}})  \right\}}_{\mathcal{D}_0}
		 +\sum_{i=1}^{f}\lambda_i\underbrace{\mathbb{E}_{\bm{s}_i} \mathbb{E}_{\tilde{\bm{y}}\sim q_{\tilde{\bm{y}}|\bm{s}_i}} \left\{-\text{log} \ p_{\bm{s}_i|\bm{\hat{x}}}(\bm{s}_i|\bm{\hat{x}})  \right\}}_{\mathcal{D}_i}+\mathcal{T},
	\end{align}
		\vspace{-5pt}
	\end{figure*}

\setcounter{equation}{\value{TempEqCnt1}}

  The goal of the  proposed variational autoencoder approach is to find the parameters  $\bm{\phi}$ and $\bm{\theta}$ that minimize the Kullback-Leibler (KL) divergences between   PDFs $p_{\tilde{\bm{y}}|\bm{x}} (\tilde{\bm{y}}|\bm{x};\bm{\phi})$ in \eqref{variational1} and  $p_{\bm{\tilde{y}}|\bm{x}}(\bm{\tilde{y}}|\bm{x};\bm{\theta})$ calculated from \eqref{equ:pos1}, and PDFs $p_{\tilde{\bm{y}}|\bm{s}_i} (\tilde{\bm{y}}|\bm{s}_i;\bm{\phi})$ in \eqref{variational2} and  $p_{\bm{\tilde{y}}|\bm{s}_i}(\bm{\tilde{y}}|\bm{s}_i;\bm{\theta})$ calculated from \eqref{equ:pos2}. Here, to avoid confusion, we use 
$q_{\tilde{\bm{y}}|\bm{s}_i}(\tilde{\bm{y}}|\bm{s}_i;\bm{\phi})$ and $q_{\tilde{\bm{y}}|\bm{x}}(\tilde{\bm{y}}|\bm{x};\bm{\phi})$ to denote  PDFs $p_{\tilde{\bm{y}}|\bm{s}_i}(\tilde{\bm{y}}|\bm{s}_i;\bm{\phi})$ and $p_{\tilde{\bm{y}}|\bm{x}}(\tilde{\bm{y}}|\bm{x};\bm{\phi})$ in \eqref{variational1} and \eqref{variational2}. For convenience, PDFs $q_{\tilde{\bm{y}}|\bm{s}_i} (\tilde{\bm{y}}|\bm{s}_i;\bm{\phi})$, $q_{\tilde{\bm{y}}|\bm{x}} (\tilde{\bm{y}}|\bm{x};\bm{\phi})$, $p_{\tilde{\bm{y}}|\bm{s}_i} (\tilde{\bm{y}}|\bm{s}_i;\bm{\theta})$, $p_{\tilde{\bm{y}}|\bm{x}} (\tilde{\bm{y}}|\bm{x};\bm{\theta})$, $p_{\bm{s}_i|\bm{\tilde{y}}}(\bm{s}_i|\bm{\tilde{y}};\bm{\theta})$,  and $p_{\bm{x}|\bm{\tilde{y}}}(\bm{x}|\bm{\tilde{y}};\bm{\theta})$ are expressed as $q_{\tilde{\bm{y}}|\bm{s}_i}$, $q_{\tilde{\bm{y}}|\bm{x}}$,  $p_{\tilde{\bm{y}}|\bm{s}_i}$, $p_{\tilde{\bm{y}}|\bm{x}}$, $p_{\bm{s}_i|\bm{\tilde{y}}}$, and $p_{\bm{x}|\bm{\tilde{y}}}$, respectively. Then, the optimization problem for parameters $\bm{\phi}$ and $\bm{\theta}$ is formulated as the minimization of the weighted sum of the  expected KL divergences over the prior PDFs $p_{\bm{x}}(\bm{x})$ and $p_{\bm{s}_i}(\bm{s}_i)$, i.e.,
\begin{align} 
	\min_{\{\bm{\theta},\bm{\phi}\}} &\  \mathbb{E}_{\bm{x}} \left\{ \mathcal{K} \left(q_{\tilde{\bm{y}}|\bm{x}}(\tilde{\bm{y}}|\bm{x};\bm{\phi})||  p_{\tilde{\bm{y}}|\bm{x}}(\tilde{\bm{y}}|\bm{x};\bm{\theta})    \right) \right\} \nonumber \\
	&+ \sum_{i=1}^{f}\lambda_i\mathbb{E}_{\bm{s}_i}\left\{ \mathcal{K} \left(q_{\tilde{\bm{y}}|\bm{s}_i}(\tilde{\bm{y}}|\bm{s}_i;\bm{\phi})||  p_{\tilde{\bm{y}}|\bm{s}_i }(\tilde{\bm{y}}|\bm{s}_i;\bm{\theta})    \right) \right\} \nonumber \\
\label{equ:1}	&\  \triangleq \mathcal{L}_{\bm{\phi},\bm{\theta}},
\end{align}
where $\lambda_i \geq 0$ and $\mathcal{K}(q(\bm{x})||p(\bm{x}))$ denotes the KL divergence between  PDFs $q(\bm{x})$ and $p(\bm{x})$. It is worth noticing that the minimization of the KL divergence is non-trivial to be solved, due to the coupled parameters $\bm{\phi}$ and $\bm{\theta}$.  Hence,  some reformulations for problem  \eqref{equ:1} is required. 

\setcounter{equation}{13}
\begin{Proposition} 
\label{pro:1}
	Function $\mathcal{L}_{\bm{\phi},\bm{\theta}}$ in \eqref{equ:1}  is upper bounded as \eqref{equ:upper}, 
	where $\mathcal{T}=\mathbb{E}_{\bm{x}}\{\log p_{\bm{x}}(\bm{x})\}+\sum_{i=1}^{f}\lambda_i\mathbb{E}_{\bm{s}_i}\{\log p_{\bm{s}_i}(\bm{s}_i)\}$ is constant and $\lambda=\sum_{i=1}^{f}\lambda_i$.
\end{Proposition}
\begin{IEEEproof}
	Please see Appendix A.
\end{IEEEproof}

\begin{Remark}
	From Proposition \ref{pro:1}, we observe that:
		\begin{itemize}
		\item The optimization problem for minimizng the objective function $\mathcal{L}_{\bm{\phi},\bm{\theta}}$  can be  relaxed by minimizing its upperbound given in \eqref{equ:upper}. By removing the constant $\mathcal{T}$, the optimization problem in \eqref{equ:1} is reformulated as 
		\begin{align} \label{equ:relax}
		\min_{\{\bm{\phi},\bm{\theta}\}}\  	\mathcal{R}+ \frac{1}{1+\lambda} \mathcal{D}_0+\sum_{i=1}^{f}\frac{\lambda_i}{1+\lambda} \mathcal{D}_i
		\triangleq \hat{\mathcal{L}}_{\bm{\phi},\bm{\theta}}.
		\end{align}
		It is worth noticing that problem \eqref{equ:relax} 
	   is also a reformulation of the rate-distortion problem in \eqref{rate-dis}, which approximates the intractable PMF $P_{\tilde{\bm{y}}}(\tilde{\bm{y}})$ with pdf $p_{\tilde{\bm{y}}}(\tilde{\bm{y}})$ and expresses the distortions in a probabilistic manner. 
	    Specifically, $\mathcal{R}$ represents the expected coding rate of  feature $\bm{\tilde{y}}$ and
		 $\mathcal{D}_0$ represents the distortion between the original data $\bm{x}$ and the recovered data $\bm{\hat{x}}$. Notably, $\mathcal{D}_i$ is the conditional entropy of the semantic information $\bm{s}_i$ given the recovered data $\bm{\hat{x}}$. The smaller the term $\mathbb{E}_{\bm{s}_i} \mathbb{E}_{\tilde{\bm{y}}\sim q_{\tilde{\bm{y}}|\bm{s}_i}} \left\{-\text{log} \ p_{\bm{s}_i|\bm{\hat{x}}}(\bm{s}_i|\bm{\hat{x}})  \right\}$ is, the more mutual information between the recovered data $\bm{\hat{x}}$ and the semantic information $\bm{s}_i$ is obtained. Since the semantic information recovery follows the optimal MAP scheme in \eqref{seman-rec}, minimizing  $\{\mathcal{D}_i\}$ is capable to reduce the distortion of the semantic information.
		 
		 		\item In optimization problem \eqref{equ:relax}, we use  weights $\{\lambda_i\}$ to balance the trade-offs between the coding rate and the distortions of both the data and the semantic information recovery. Specifically,   when  $\lambda_1=\lambda_2=\cdots=\lambda_f=0$,  problem \eqref{equ:relax} equals to the conventional data compression problem \cite{Balle2017,Balle2018,9242247} without considering the semantic information.  
		 		
%		Specifically, with the increasing of $\lambda$, the weights for both the distortion of the semantic information and the coding rate in problem \eqref{equ:relax} will increase, while the distortion of the image recovery will be disregarded. However, in the rate-distortion problem \eqref{equ:ratedistortion}, with the increasing of $\alpha_3$, both the distortion of the image recovery and the coding rate will  be disregarded. 
		\end{itemize}
\end{Remark}

\subsection{Rate-distortion optimization for general data distribution}
This subsection introduces the computations of the  proposed rate-distortion problem in \eqref{equ:relax} for general data distributions. 
For the distortion $\mathcal{D}_0$ given in \eqref{equ:relax},  distribution $p_{\bm{x}|\bm{\hat{x}}}(\bm{x}|\bm{\hat{x}})$ can be approximated by  a distribution of the exponential family \cite{Balle2017,markov}, i.e.,
\begin{align}
		p_{\bm{x}|\bm{\hat{x}}}(\bm{x}|\bm{\hat{x}}) = \frac{1}{Z(\alpha)} \text{exp}\left(-\alpha d(\bm{\hat{x}},\bm{x})   \right),
\end{align}
where $\alpha>0$, $Z(\alpha)$ is a constant to normalize  PDF $p_{\bm{x}|\bm{\hat{x}}}$, and function $d(\bm{\hat{x}},\bm{x})$ is the  distortion measure, in terms of the distance between the recovered data $\hat{\bm{x}}$ and the original data $\bm{x}$. For image data, $p_{\bm{x}|\bm{\hat{x}}}(\bm{x}|\bm{\hat{x}})=\mathcal{N}(\bm{x}|\bm{\hat{x}},(2\alpha)^{-1}\bm{I})$ with $\alpha>0$ is  used \cite{Balle2017} and it  corresponds to the most widely-adopted MSE distortion, i.e., $d(\bm{\hat{x}},\bm{x})=||\bm{x}-\bm{\hat{x}}||^2$. For text data, BLEU \cite{jiang2022deep} is usually used as the distortion function $d(\bm{\hat{x}},\bm{x})$.

The computation of  $\mathcal{D}_i$ in \eqref{equ:relax} depends on the semantic task. However, for most of semantic information, e.g., image's label information, the true PDF $p_{\bm{s}_i|\bm{\hat{x}}}(\bm{s}_i|\bm{\hat{x}})$ is difficult to be obtained. An alternative approach is to use a parameterized density function $Q_{\bm{s}_i|\bm{\hat{x}}}(\bm{s}_i|\bm{\hat{x}};\bm{\gamma})$ with parameter $\bm{\gamma}$ to approximate the true PDF $p_{\bm{s}_i|\bm{\hat{x}}}(\bm{s}_i|\bm{\hat{x}})$ \cite{hossain2019comprehensive,krizhevsky2017imagenet}. 
Then, we have the following result.
\begin{Lemma} \label{lemma1}
By replacing the true PDF $p_{\bm{s}_i|\bm{\hat{x}}}(\bm{s}_i|\bm{\hat{x}})$ in \eqref{equ:relax} with the approximated density function $Q_{\bm{s}_i|\bm{\hat{x}}}(\bm{s}_i|\bm{\hat{x}};\bm{\gamma})$, function $\hat{\mathcal{L}}_{\bm{\phi},\bm{\theta}}$ given in \eqref{equ:relax} is upper bounded as 
\begin{align} \label{lemma:equ1}
		\hat{\mathcal{L}}_{\bm{\phi},\bm{\theta}} &\leq \mathcal{R}+\frac{1}{1+\lambda}\mathcal{D}_0 \nonumber \\
		&\ \ +\sum_{i=1}^{f}\frac{\lambda_i}{1+\lambda} \underbrace{\mathbb{E}_{\bm{s}_i} \mathbb{E}_{\tilde{\bm{y}}\sim q_{\tilde{\bm{y}}|\bm{s}_i}} \left\{-\text{log} \ Q_{\bm{s}_i|\bm{\hat{x}}}(\bm{s}_i|\bm{\hat{x}};\bm{\gamma})  \right\}}_{\hat{\mathcal{D}}_i}\nonumber \\
		&\ \ \triangleq \hat{\mathcal{L}}_{\bm{\phi},\bm{\theta},\bm{\gamma}}.
\end{align}
\end{Lemma}
\begin{IEEEproof}
It can be easily checked that
\begin{align} \label{hatD}
	\hat{\mathcal{D}}_i=\mathcal{D}_i+\mathbb{E}_{\bm{s}_i}\left\{ \mathcal{K}\left\{ p_{\bm{s}_i|\bm{\hat{x}}}(\bm{s}_i|\bm{\hat{x}})||Q_{\bm{s}_i|\bm{\hat{x}}}(\bm{s}_i|\bm{\hat{x}};\bm{\gamma}) \right\}\right\},
\end{align}which implies $\hat{\mathcal{L}}_{\bm{\phi},\bm{\theta},\bm{\gamma}}\geq \hat{\mathcal{L}}_{\bm{\phi},\bm{\theta}}$ due to $\mathcal{K}\left\{ p_{\bm{s}_i|\bm{\hat{x}}}(\bm{s}|\bm{\hat{x}})||Q_{\bm{s}_i|\bm{\hat{x}}}(\bm{s}|\bm{\hat{x}};\bm{\gamma}) \right\}\geq 0$. 
\end{IEEEproof}

It is worth noticing that    $\hat{\mathcal{D}}_i$ in \eqref{lemma:equ1}
is the classical cross entropy, which is widely used in  image classification and recognition  tasks \cite{he2016deep,krizhevsky2017imagenet}. By minimizing  $\hat{\mathcal{D}}_i$ in \eqref{lemma:equ1}, the KL divergence between the true distribution $p_{\bm{s}_i|\bm{\hat{x}}}(\bm{s}_i|\bm{\hat{x}})$ and the approximated density function $Q_{\bm{s}_i|\bm{\hat{x}}}(\bm{s}_i|\bm{\hat{x}};\bm{\gamma})$ will be reduced.

In a summary, the approximated rate-distortion optimization problem for the JTD-SC problem is given as 
\begin{align} \label{problem:1}
	\min_{\{\bm{\phi},\bm{\theta},\bm{\gamma}\}}  \hat{\mathcal{L}}_{\bm{\phi},\bm{\theta},\bm{\gamma}}.
\end{align}
Minimizing the objective function of this problem not only results in resolving the rate-distortion problem in \eqref{equ:relax}, but also fits the approximated density function $Q_{\bm{s}_i|\bm{\hat{x}}}(\bm{s}_i|\bm{\hat{x}};\bm{\gamma}), i\in[f],$ into the true PDF $p_{\bm{s}_i|\bm{\hat{x}}}(\bm{s}_i|\bm{\hat{x}})$.

Now, a big challenge for solving problem \eqref{problem:1} is how to characterize the PDF of  random variable $\bm{\tilde{y}}$, which is calculated from  $p_{{\bm{y}}}({\bm{y}})$ according to \eqref{hyper:1}. To characterize  $p_{{\bm{y}}}({\bm{y}})$, a large number of existing works have made different assumptions and constraints on random variable $\bm{y}$. A suboptimal while simple assumption is sparse prior \cite{fastsparse}, i.e., $\bm{y}$ follows sparse distribution \cite{sparseprior},  and then the rate $-\log p_{\tilde{\bm{y}}}(\tilde{\bm{y}})$ in \eqref{equ:upper} can be relaxed by a  regularization term related to norm-$1$ $||\bm{y}||_{1}$ \cite{fastsparse,sparseprior}. However, the sparse assumption imposes  too strong limit on the distribution of  feature $\bm{y}$. In a more flexible way, the authors in \cite{Balle2017}  approximate the true PDF $p_{y_i}(y_i)$ with a  non-parametric fully-factorized density model, i.e., $q(y_i|\varphi)$, where $\varphi$ is the parameter needed to be learned. However, both  the fixed prior hypothesis and the learning-based prior model ignore the dependencies among the elements of feature vector $\bm{y}$, possibly resulting in higher coding rate and distortions for the JTD-SC problem.

\section{Case Study: Image Transmission and Classification }
This section takes  image transmission and classification task as examples to validate the proposed variational autoencoder approach for the JTD-SC problem. 
For the image data, there exists significant spatial dependencies among the elements of the image feature $\bm{y}$ \cite{Balle2018}, and more importantly, the spatial dependencies vary across the images with the different semantic information. An efficient solution is to estimate the density model of each feature vector and then transmit it together with the feature vector, which is known as the FA scheme \cite{9242247}. To accurately model  PDF $p_{\bm{{y}}}(\bm{{y}})$ and facilitate a better compression performance, we 
combine the well-known FA scheme with the proposed variational autoencoder, and propose an FA-based autoencoder scheme for  image transmission and classification.

In this section,  we first introduce the proposed framework of the FA-based autoencoder scheme; then, we model the transmission framework as a Bayesian model similar to Fig. \ref{Appendix:2} and derive the new optimization problem; Next, we present the detailed module design of the FA-based scheme; Finally, we propose an iterative training algorithm to train the parameters.
 \begin{figure*}[t]
\normalsize
\setlength{\abovecaptionskip}{+0.3cm}
\setlength{\belowcaptionskip}{-0.1cm}
	\centering
	\includegraphics[width=6.0in]{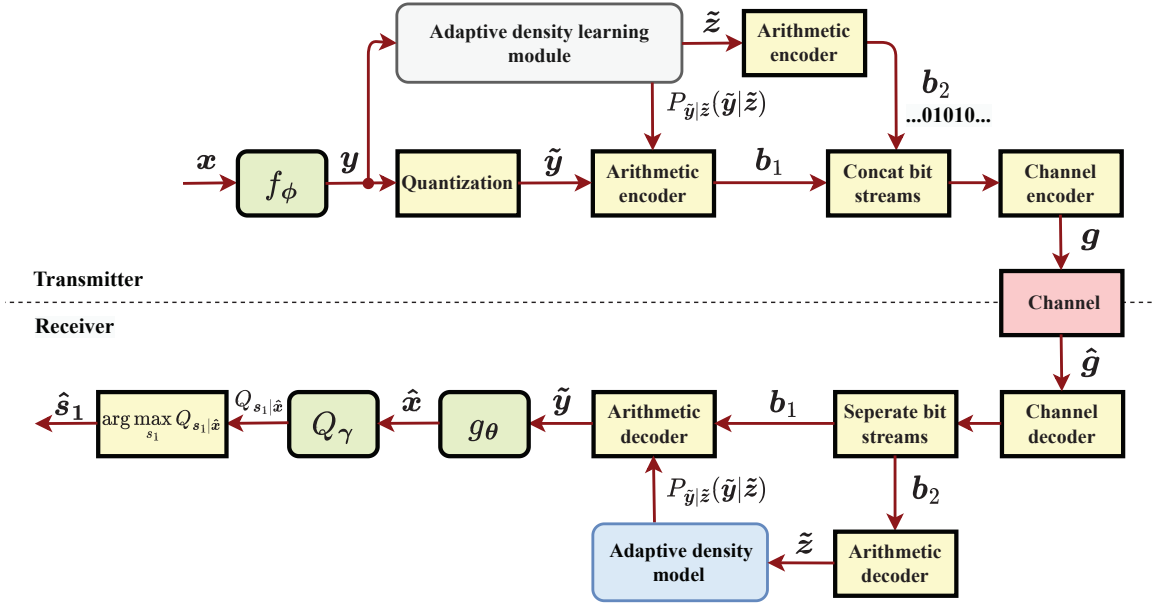}
	\caption{Framework of the FA-based autoencoder method}
	\label{Appendix:4}
\end{figure*}
\subsection{FA-based autoencoder scheme}
In this subsection, we present the  architecture of the FA-based autoencoder  scheme for  image transmission and classification, as shown in Fig. \ref{Appendix:4}. Here, image's label information is regarded as  the semantic information $\bm{s}_1$ that needs to be recovered.  Compared with the general framework as shown in Fig. \ref{figure:one}, the FA-based autoencoder scheme introduces the density learning module at the transmitter side to learn the PDF of each feature vector $\bm{\tilde{y}}$, and brings in the adaptive density  model at the receiver side to assist in decoding feature vector $\bm{\tilde{y}}$ from bit streams. By  characterizing the PDF of feature  $\bm{y}$, the FA-based method can adaptively construct the codebook for the encoding and decoding of the feature vector, which results in higher coding gain. In the followings, we first present the design of the density learning module by following the nonlinear transformer coding method \cite{9242247}, and then the frameworks of the transmitter and the receiver are introduced.

\begin{enumerate}
	\item Adaptive density learning module: A standard way to model the dependencies among the elements of the feature vector is to introduce latent variables conditioned on which the elements are assumed to be independent \cite{Balle2018}. 
 Here, we introduce an additional set of random variables $\bm{\tilde{z}}=[\tilde{z}_1,\tilde{z}_2,\cdots,\tilde{z}_{D}]^T\in \mathbb{R}^{D}$ to capture the spatial dependencies of $\bm{y}$. The distribution of  $y_i$ conditioned on  $\bm{\tilde{z}}$ can be modeled as the independent while not identically distributed Gaussian random variable  with zero mean and  variance $\sigma_{i}^2$ \cite{Balle2018}, i.e.,
 \begin{align} \label{norm:1}
 	p_{y_i|\bm{z}}(y_i|\bm{\tilde{z}})=\mathcal{N}(y_i;0,\sigma_i^2),
 \end{align}
 where  $\sigma_{i}$ is estimated by applying a transform function $h_{\psi_2}$ with parameter $\psi_2$ to $\bm{\tilde{z}}$, i.e., $\bm{\sigma}=h_{\psi_2}(\bm{\tilde{z}})$ with $\bm{\sigma}=[\sigma_1,\sigma_2,\cdots,\sigma_{K}]^T$.   According to  
 the Bayesian expression in \eqref{hyper:1} and \eqref{norm:1},   PMF $P_{\tilde{\bm{y}}|\bm{\tilde{z}}}(\tilde{\bm{y}}|\bm{\tilde{z}})$ is calculated as
 \begin{align} \label{den:y}
 		P_{\bm{\tilde{y}}|\bm{\tilde{z}}}(\bm{\tilde{y}}=\bm{k}|\bm{\tilde{z}})= \prod_{i=1}^{K} \left\{\int_{k_i-0.5}^{k_i+0.5}\mathcal{N}(y_i;0,\sigma_i^2)dy_i\right\}, 
 \end{align}
with $\bm{k}=[k_1,k_2,\cdots,k_K]^T\in \mathbb{Z}^{K}$. It is worth noticing that the latent variable $\bm{\tilde{z}}$ is  often referred as the \emph{side information} of  feature $\bm{y}$ \cite{Balle2018}, and needs to be estimated at the transmitter and sent to the receiver for decoding feature $\bm{y}$ from the received bit streams. As shown in Fig. \ref{Appendix:5}, we involve  function $h_{\bm{\psi}_1}$ with parameter $\bm{\psi}_1$ to transform feature vector $\bm{y}$ to vector $\bm{z}\in \mathbb{R}^{D}$.  PDF $p_{\tilde{z}}(\bm{z})$ can be approximated by using the non-parametric fully-factorized density model \cite{Balle2018}, i.e.,
 \begin{align} \label{factor:1}
 	p_{\bm{z}}(\bm{z};\bm{\omega})=\prod_{i=1}^{D}p_{z_i}(z_{i};\omega_i)
 \end{align}
 where $\omega_i$ is the parameter used to  characterize the density function $p_{z_i}(z_{i};\omega_i)$ and $\bm{\omega}=\{\omega_1,\omega_2,\cdots,\omega_D\}$. $\bm{\tilde{z}}$ is the quantized version of vector $\bm{z}$, whose PMF is calculated as 
 \begin{align} \label{den:z}
 	 		P_{\bm{\tilde{z}}}(\bm{\tilde{z}}=\bm{k};\bm{\omega})&= \prod_{i=1}^{D} \left\{\int_{k_i-0.5}^{k_i+0.5}p_{z_i}(z_{i};\omega_i)dz_i\right\},  \end{align}
 with $\bm{k}=[k_1,k_2,\cdots,k_D]^T\in \mathbb{Z}^{D}$.
  The quantized vector $\bm{\tilde{z}}$ is transmitted to the receiver to construct PMF $P_{\tilde{y}_i|\bm{\tilde{z}}}(\tilde{y}_i|\bm{\tilde{z}})$ according to the adaptive density model in \eqref{den:y}. 

   \begin{figure*}[t]
\normalsize
\setlength{\abovecaptionskip}{+0.3cm}
\setlength{\belowcaptionskip}{-0.1cm}
	\centering
	\includegraphics[width=5.1in]{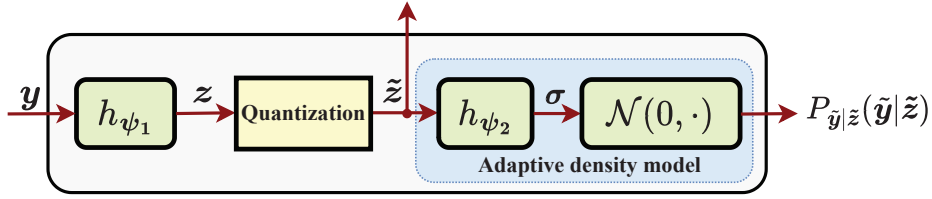}
	\caption{Adaptive density learning module}
	\label{Appendix:5}
\end{figure*}

  \item Transmitter design: First,  image $\bm{x}$ is fed into the parametric function $f_{\bm{\phi}}(\bm{x})$ at the transmitter to extract feature $\bm{y}$ of the image. Then, the extracted feature $\bm{y}$ is input into two operation queues: On the one hand, $\bm{y}$ is quantized as $\bm{\tilde{y}}$ which waits for arithmetic encoding; on the other hand, it is fed into  function $h_{\bm{\psi}_1}(\bm{y})$ to generate its latent feature $\bm{z}$. Next, the obtained latent feature $\bm{z}$ is quantized and fed into  transform function $h_{\bm{\psi}_2}(\bm{\tilde{z}})$ to estimate the standard deviations $\bm{\sigma}$, i.e., $\bm{\sigma}=h_{\bm{\psi}_2}(\bm{\tilde{z}})$.   Based on the computed PMF $P_{\bm{\tilde{y}}|\bm{\tilde{z}}}(\bm{\tilde{y}}|\bm{\tilde{z}})$ given in \eqref{den:y}, arithmetic encoder \cite{gallager1968information}  is adopted to encode the quantized feature $\bm{\tilde{y}}$ into bit streams. The latent feature $\bm{\tilde{z}}$ is also encoded into bit streams by  arithmetic encoder according to the density model given in \eqref{den:z}. Finally, the two flows of the bit streams are concatenated and encoded as symbols $\bm{g}$ by utilizing certain channel coding, and are transmitted to the receiver over the error-free channel. It is worth noting that the sizes of the two bitstreams also need to be transmitted to the receiver to help the receiver separate the concatenated bitstreams. However, the bits used to encode the sizes of the bitstream are far less than those for encoding the image features, and thus their effects on the coding rate are ignored.
 
  \item Receiver design: After receiving from the transmitter, the receiver first decodes the transmitted bits from the corrupted symbols $\bm{\hat{g}}$ through channel decoder. The decoded bits are separated into two parts: First, the latent feature $\bm{\tilde{z}}$ is decoded by applying the arithmetic decoder according to the density model given in \eqref{den:z}. Then, $\bm{\tilde{z}}$ is fed into function $h_{\bm{\psi}_2}$, which has the same parameter $\bm{\psi}_2$ as the one at the transmitter,  to compute PMF $P_{\bm{\tilde{y}}|\bm{\tilde{z}}}(\bm{\tilde{y}}|\bm{\tilde{z}})$. With the obtained PMF $P_{\bm{\tilde{y}}|\bm{\tilde{z}}}(\bm{\tilde{y}}|\bm{\tilde{z}})$ and the decoded bit stream, we decode  feature $\bm{\tilde{y}}$ by utilizing the arithmetic decoder. Next, the decoded feature $\bm{\tilde{y}}$ is fed into function $g_{\bm{\theta}}(\bm{\tilde{y}})$ to recover the image, i.e., $\bm{\hat{x}}=g_{\bm{\theta}}(\bm{\tilde{y}})$. Then, the recovered image $\bm{\hat{x}}$ is input into the function $Q_{\bm{\gamma}}$ and the output is the approximated probability $Q_{\bm{s}_1|\bm{\hat{x}}}(\bm{s}_1|\bm{\hat{x}};\bm{\gamma})$. According to the adopted MAP detection scheme, the detected label information $\hat{\bm{s}}_1$ is obtained as
  \begin{align}
  	\bm{\hat{s}}_1= \arg \max_{\bm{s}_1\in \mathbb{S}_1} Q_{\bm{s}_1|\bm{\hat{x}}}(\bm{s}_1|\bm{\hat{x}};\bm{\gamma}).
  \end{align}
  
\end{enumerate}

\subsection{Bayesian model with latent variables $\bm{z}$ and $\bm{\tilde{z}}$ }
\begin{figure}[htbp]
\centering
\subfigure[Inference model]{
\begin{minipage}[t]{1\linewidth}
	\centering
	\includegraphics[width=2.6in]{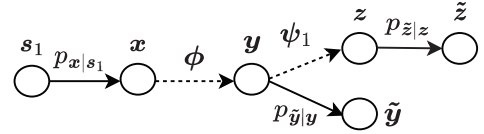}
%\caption{fig1}
\end{minipage}%
}%

\subfigure[Generative model]{
\begin{minipage}[t]{1\linewidth}
	\centering
	\includegraphics[width=2.2in]{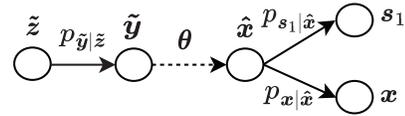}
%\caption{fig2}
\end{minipage}%
}%
\centering
\caption{Extended Bayesian model with the latent variables}
\label{Appendix:3}
\end{figure}

\setcounter{TempEqCnt1}{\value{equation}}
\setcounter{TempEqCnt}{\value{equation}}
\begin{figure*}[b]
		  \hrulefill
		\vspace{-5pt}
		\setcounter{equation}{29}
\begin{align} \label{loss:final}
		\min_{\{\bm{\phi,\psi_1,\psi_2,\theta},\bm{\gamma},\bm{\omega}\}} \ 	 &\frac{1}{1+\lambda_1}\bar{\mathcal{D}}_0 + \frac{\lambda_1}{1+\lambda_1}\bar{\mathcal{D}}_1 + \mathbb{E}_{\bm{x}} \mathbb{E}_{\tilde{\bm{y}},\tilde{\bm{z}}\sim q_{\tilde{\bm{y}},\tilde{\bm{z}}|\bm{x}}}\left\{-\log p_{\tilde{\bm{y}}|\bm{\tilde{z}}}(\tilde{\bm{y}}|\bm{\tilde{z}})-\log p_{\tilde{\bm{z}}}(\tilde{\bm{z}};\bm{\omega})\right\}.
\end{align}
		\vspace{-5pt}
		\setcounter{TempEqCnt}{\value{equation}}
\end{figure*}
\setcounter{equation}{\value{TempEqCnt1}}
According to the FA-based transmission scheme shown in Fig. \ref{Appendix:4}, we extend the Bayesian model in Fig. \ref{Appendix:2}  into Fig. \ref{Appendix:3} by adding the latent variables $\bm{z}$ and $\bm{\tilde{z}}$. In order to use the gradient descent methods to optimize the trainable parameters $\{\bm{\phi},\bm{\psi}_1,\bm{\psi}_2,\bm{\theta},\bm{\gamma}$, $\bm{\omega}\}$, the errors caused by quantizing  variables $\bm{z}$ and $\bm{y}$ are approximated by the i.i.d. uniform noise \cite{Balle2017}. Therefore, PDFs $p_{\tilde{\bm{y}},\bm{\tilde{z}}|\bm{s}_1}(\tilde{\bm{y}},\bm{\tilde{z}}|\bm{s}_1;\bm{\phi},\bm{\psi_1})$ and $p_{\tilde{\bm{y}},\bm{\tilde{z}}|\bm{x}}(\tilde{\bm{y}},\bm{\tilde{z}}|\bm{x};\bm{\phi},\bm{\psi_1})$, denoted by $q_{\tilde{\bm{y}},\bm{\tilde{z}}|\bm{s}_1}(\tilde{\bm{y}},\bm{\tilde{z}}|\bm{s}_1;\bm{\phi},\bm{\psi_1})$ and $q_{\tilde{\bm{y}},\bm{\tilde{z}}|\bm{x}}(\tilde{\bm{y}},\bm{\tilde{z}}|\bm{x};\bm{\phi},\bm{\psi_1})$, are calculated as
\begin{small}
\begin{align}
	\label{variational1_2}	q_{\bm{\tilde{y}},\bm{\tilde{z}}|\bm{{x}}}(\bm{\tilde{y}},\bm{\tilde{z}}|\bm{{x}};\bm{\phi},\bm{\psi_1})&=\left( \prod \limits_{i=1}^{K}\mathcal{U}(\tilde{y}_i;y_i,1)\right) \left(\prod \limits_{i=1}^{D}\mathcal{U}(\tilde{z}_i;z_i,1)\right),\\
\label{variational2_2}	q_{\bm{\tilde{y}},\bm{\tilde{z}}|\bm{s}_1}(\bm{\tilde{y}},\bm{\tilde{z}}|\bm{s}_1;\bm{\phi},\bm{\psi_1})&=\int_{\bm{x}} q_{\bm{\tilde{y}},\bm{\tilde{z}}|\bm{{x}}}(\bm{\tilde{y}},\bm{\tilde{z}}|\bm{{x}};\bm{\phi},\bm{\psi_1})p_{\bm{x}|\bm{s}_1}(\bm{x}|\bm{s}_1) d \bm{x}, 
\end{align}
\end{small}with $\bm{y}=f_{\bm{\phi}}(\bm{x})$ and $\bm{z}=h_{\bm{\psi_1}}({{\bm{y}}})$.
According to \eqref{norm:1}, PDF $p_{\bm{\tilde{y}}|\bm{\tilde{z}}}(\bm{\tilde{y}}|\bm{\tilde{z}})$ is given as 
 \begin{align} 
 		p_{\bm{\tilde{y}}|\bm{\tilde{z}}}(\bm{\tilde{y}}|\bm{\tilde{z}})&= \prod_{i=1}^{K} \left\{\mathcal{N}(y_i;0,\sigma_i^2)*\mathcal{U}(\tilde{y}_i;y_i,1)\right\}, \\
 		&=\prod_{i=1}^{K} \left\{\int_{\tilde{y}_i-0.5}^{\tilde{y}_i+0.5} \mathcal{N}(y_i;0,\sigma_i^2) dy_i\right\},\\
 \label{condition:3}		&=\prod_{i=1}^{K} \left\{\mathcal{C}(\tilde{y}_i+0.5;0,\sigma_i^2)- \mathcal{C}(\tilde{y}_i-0.5;0,\sigma_i^2)\right\},
 \end{align}
 where $\mathcal{C}(\tilde{y}_i;0,\sigma_i^2)$ is the cumulative function of $\mathcal{N}(\tilde{y}_i;0,\sigma_i^2)$.  Similarly,  according to \eqref{factor:1}, PDF $p_{\bm{\tilde{z}}}(\bm{\tilde{z}};\bm{\omega})$ is given as 
 \begin{align}
 	p_{\bm{\tilde{z}}}(\bm{\tilde{z}};\bm{\omega})=\prod_{i=1}^{D}\left\{p_{z_{i}}(z_i;\omega_i)*\mathcal{U}(\tilde{z}_i;z_i,1)\right\}
 \end{align}

Given the extended Bayesian model in Fig. \ref{Appendix:3},  the variational autoencoder approach  approximates PDFs $q_{\tilde{\bm{y}},\bm{\tilde{z}}|\bm{s}_1}(\tilde{\bm{y}},\bm{\tilde{z}}|\bm{s}_1;\bm{\phi},\bm{\psi_1})$ and $q_{\tilde{\bm{y}},\bm{\tilde{z}}|\bm{x}}(\tilde{\bm{y}},\bm{\tilde{z}}|\bm{x};\bm{\phi},\bm{\psi_1})$ in \eqref{variational1_2} and \eqref{variational2_2} as PDFs $p_{\bm{\tilde{y},\bm{\tilde{z}}}|\bm{s}_1}(\bm{\tilde{y},\bm{\tilde{z}}}|\bm{s}_1; \bm{\theta},\bm{\psi_2})$ and $p_{\bm{\tilde{y},\bm{\tilde{z}}}|\bm{x}}(\bm{\tilde{y},\bm{\tilde{z}}}|\bm{x};\bm{\theta},\bm{\psi_2})$, respectively. By following Proposition \ref{pro:1} and Lemma \ref{lemma1},  the rate-distortion problem  for the joint  image transmission and classification  can be easily obtained by replacing $\bm{\tilde{y}}$ with the pair of  random variables $(\bm{\tilde{y}},\bm{\tilde{z}})$, which is given by \eqref{loss:final}. In  problem \eqref{loss:final}, $\bar{\mathcal{D}}_0$ and $\bar{\mathcal{D}}_1$ are obtained from $\mathcal{D}_0$ in \eqref{equ:upper} and $\hat{\mathcal{D}}_1$ in \eqref{hatD}, respectively, by replacing $\bm{\tilde{y}}$ with the pair of  random variables $(\bm{\tilde{y}},\bm{\tilde{z}})$,  $-\log p_{\tilde{\bm{y}}|\bm{\tilde{z}}}(\tilde{\bm{y}}|\bm{\tilde{z}})$ is the coding rate of feature $\tilde{\bm{y}}$ given the latent variable $\bm{\tilde{z}}$, and $-\log p_{\tilde{\bm{z}}}(\bm{\tilde{z}};\bm{\omega})$ is the coding rate of the latent variable $\tilde{\bm{z}}$.

\subsection{Implementation issues}
In this subsection, we present the implementation details of each module in the proposed FA-based autoencoder scheme by applying the DNN architecture. As shown in Fig. \ref{Appendix:4} and \ref{Appendix:5}, the proposed scheme mainly consists of 
the feature extraction function $f_{\bm{\phi}}$, the image recovery function $g_{\bm{\theta}}$, the adaptive density functions $h_{\bm{\psi}_1}$ and $h_{\bm{\psi}_2}$, and the semantic recovery function  $Q_{\bm{\gamma}}$, which are described as follows:
  \begin{figure*}[t]
\normalsize
\setlength{\abovecaptionskip}{+0.3cm}
\setlength{\belowcaptionskip}{-0.1cm}
	\centering
	\includegraphics[width=5.1in]{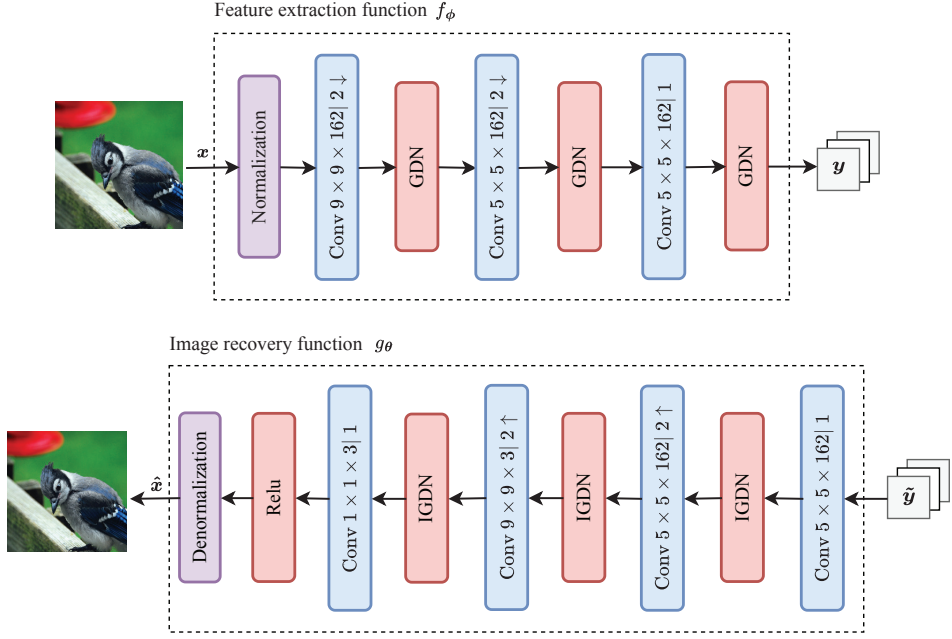}
	\caption{Network architectures of functions $f_{\bm{\phi}}$ and $g_{\bm{\theta}}$. The parameters of the CNNs are described as follows:  kernel support height $\times$ kernel support width $\times$ number of filters / down or upsampling stride, where  $\uparrow$ and $\downarrow$ denote upsampling and  downsampling, respectively. }
	\label{N1}
\end{figure*}

\begin{enumerate}
  \item Functions $f_{\bm{\phi}}$ and $g_{\bm{\theta}}$: The key of designing the feature extraction function $f_{\bm{\phi}}$ is to construct the DNN-based architecture that  compresses the input image into a low-dimensional space while maintaining the important features. To this end, we adopt the convolutional neural network (CNN) to efficiently downsample the high-dimensional images.  As shown in Fig. \ref{N1}, the image source $\bm{x} \in \mathbb{R}^{H \times W \times 3}$  with the maximal pixel value $255$ is first normalized and then fed into a sequence of CNNs for downsampling.  For the design of the CNNs, we first choose a large kernel size, i.e., $9\times9$, to extract the important features of the objects in  the high-dimensional images, and then  small kernel size, i.e., $5\times 5$, is utilized to reduce the computation complexity. The generalized divisive normalization (GDN) function \cite{balle2015density, Balle2017}, which is able to efficiently Gaussianize the feature vectors,  is chosen as the activation function of each CNN.
  
  		Function $g_{\bm{\theta}}$ inverts the compression operations performed by function $f_{\bm{\phi}}$. As shown in Fig. \ref{N1},  feature vector $\bm{\tilde{y}}$ received at the receiver will be fed into the CNN-based structure which upsamples it into the corrupted image $\bm{\hat{x}}$. The hyperparameters of the CNNs are described in Fig. \ref{N1}. Inverse GDN (IGDN) function \cite{balle2015density, Balle2017} as the inverse transform of the GDN function is chosen as the activation function of the first three CNN layers.  Notably, we add one CNN layer with Relu function at the last convolution layer  in order to transform the output into positive value. Finally, the output image is denormalized into the range $[0,255]$. 
\item The adaptive density functions $h_{\bm{\psi}_1}$ and $h_{\bm{\psi}_2}$: The overall architectures of the adaptive density functions $h_{\bm{\psi}_1}$ and $h_{\bm{\psi}_2}$ are illustrated in Fig. \ref{N2}. In function $h_{\bm{\psi}_1}$,  the CNN layers  with the GDN activation function are applied to compress   vector $\bm{y}$ into the latent variable $\bm{z}$. In function $h_{\bm{\psi}_2}$, the CNN layers  with the IGDN activation function are adopted to transform the latent variable $\bm{\tilde{z}}$ into the standard deviation  $\bm{\sigma}$.
  \begin{figure*}[t]
\normalsize
\setlength{\abovecaptionskip}{+0.3cm}
\setlength{\belowcaptionskip}{-0.1cm}
	\centering
	\includegraphics[width=5.5in]{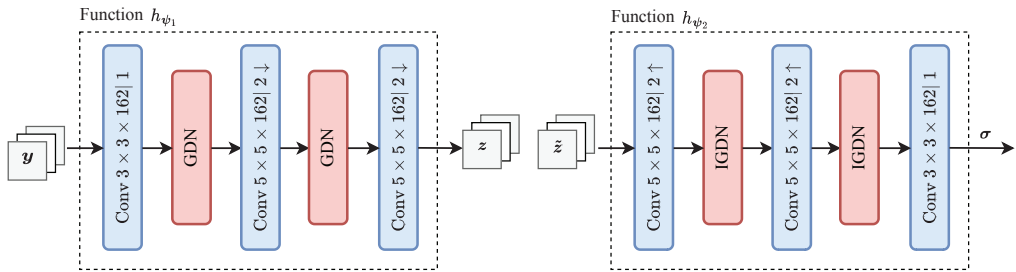}
	\caption{Network architectures of functions $h_{\psi_1}$ and $h_{\psi_2}$}
	\label{N2}
\end{figure*}
\item The semantic recovery function $Q_{\bm{\gamma}}$: Here, we adopt the widely-known Resnet-$p$ \cite{he2016deep} as the semantic recovery function $Q_{\bm{\gamma}}$ to work out the image classification task, where $p$ is the number of the residual layers in Resnet.  It is worth noticing that the Resnet-$p$ might not be the optimal architecture to approximate the true probability $p_{\bm{s}_1|\hat{\bm{x}}}$ for the arbitrary image dataset. More advanced DNN architectures can be explored to improve the classification performance.
\end{enumerate}

\subsection{Iterative training algorithm}

\setcounter{equation}{\value{TempEqCnt}}
This subsection presents the training algorithm for optimizing  the parameters $\{\bm{\phi},\bm{\psi}_1,\bm{\psi}_2,\bm{\theta},\bm{\gamma}$, $\bm{\omega}\}$ of the neural networks and aims to minimize the objective function given in \eqref{loss:final}. By approximating the expectations in \eqref{loss:final} by averaging over a set of training samples, the loss function for training the parameters is calculated as 
\begin{small}
\begin{align} \label{loss_app}
	\tilde{\mathcal{L}} & = \frac{1}{B_1}\sum_{i=1}^{B_1}\left\{\frac{1}{B_n}\sum_{j=1}^{B_n}-\log p_{\tilde{\bm{y}}|\bm{\tilde{z}}}(\tilde{\bm{y}}_{ij}|\bm{\tilde{z}}_{ij})-\log p_{\tilde{\bm{z}}}(\tilde{\bm{z}}_{ij};\bm{\omega})\right\}\nonumber \\
	&\quad \ +
	\frac{1}{1+\lambda_1}\frac{1}{B_1}\sum_{i=1}^{B_1}  \left\{ \frac{1}{B_n}\sum_{j=1}^{B_n} -\text{log} \ p_{\bm{x}|\bm{\hat{x}}}(\bm{x}_i|\bm{\hat{x}}_{ij})  \right\} \nonumber \\
	&\quad \ + \frac{\lambda_1}{1+\lambda_1} \frac{1}{B_2} \sum_{i=1}^{B_2} \left\{ \frac{1}{B_n}\sum_{j=1}^{B_n} -\text{log} \ Q_{\bm{s}_1|\bm{\hat{x}}}(\bm{s}_{1,i}|\bm{\hat{x}}_{ij};\bm{\gamma})  \right\} 
		,
\end{align}
\end{small}where $\bm{x}_i\in{\mathcal{X}}$ and $\bm{s}_{1,i}\in{\mathcal{S}}$ are the image and the semantic  training samples, respectively, $\mathcal{X}$ and $ \mathcal{S}$ are the data set of the image and the semantic information, respectively, $\hat{\bm{x}}_{ij}=g_{\theta_1}(f_{\bm{\phi}}(\bm{x}_i)+o_{j})$, $\bm{\tilde{y}}_{ij}=f_{\bm{\phi}}(\bm{x}_i)+o_j$, $\bm{\tilde{z}}_{ij}=h_{\psi_1}(f_{\bm{\phi}}(\bm{x}_i))+\hat{o}_{j}$, with $o_j$ and $\hat{o}_j$ being the $j$-th uniform noise samples, and $B_1$, $B_2$, and $B_n$ are the  mini batch sizes of the image, semantic, and noise samples, respectively.  When $B_1$, $B_2$, and $B_n$ are large enough, the loss function $\tilde{\mathcal{L}}$ approximates to the exact objective function in \eqref{loss:final} according to the  law of large numbers \cite{markov}. 

With the loss function given in \eqref{loss_app},  a straightforward way to optimize the parameters $\{\bm{\phi,\theta,\gamma}$, $\bm{\psi_1,\psi_2,\omega}\}$ is the end-to-end training method \cite{Balle2017,luo2018deepsic}, where the parameters are jointly trained epoch by epoch to minimize the loss function $\tilde{\mathcal{L}}$  by utilizing the back propagation algorithm \cite{8054694}. However, this method requires the same batch size, i.e. $B_1=B_2$, to update the parameters, which easily leads to an imbalance between the image generation and the semantic restoration. Especially for the high-dimensional images with the complex semantic information, if the batch size is too small,  the distribution of semantic information cannot be well approximated, resulting in overfitting problem \cite{dietterich1995overfitting}. If the batch size is too large, it may cause the memory explosion  and significantly slow down the training speed without improving the performance of the image generation. To tackle these problems, we propose an iterative training algorithm to efficiently train the parameters of the neural networks step by step.  Fig. \ref{N3} shows the signal flows of the proposed FA-based transmission scheme by disregarding the error caused by wireless transmission and replacing the quantization error with the uniform noise, and the gradient flows for updating the parameters can be divided into two steps:
  \begin{figure*}[t]
\normalsize
\setlength{\abovecaptionskip}{+0.3cm}
\setlength{\belowcaptionskip}{-0.1cm}
	\centering
	\includegraphics[width=6.0in]{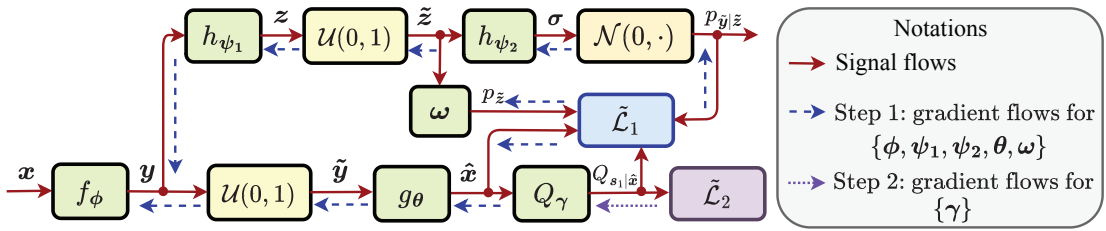}
	\caption{Signal and gradient flows for the proposed iterative training algorithm. $\mathcal{U}(0,1)$ denotes the uniform noise.}
	\label{N3}
\end{figure*}

\begin{itemize}
	\item \textbf{Step 1}: With fixed parameter $\bm{\gamma}$, train the parameters $\{\bm{\phi,\psi_1,\psi_2,\theta},\bm{\omega}\}$.   At this step, we fix the semantic recovery function $Q_{\bm{\gamma}}$ and then train the parameters $\{\bm{\phi,\psi_1,\psi_2,\theta},\bm{\omega}\}$  for $N_1>0$ epochs by utilizing the end-to-end training method.  The loss function $\tilde{\mathcal{L}}_1$ and its gradient $\nabla \tilde{\mathcal{L}}_1$ with respect to (w.r.t.) the parameters $\{\bm{\phi,\psi_1,\psi_2,\theta},\bm{\omega}\}$ are calculated according to the loss function $\tilde{\mathcal{L}}$ in \eqref{loss_app} by letting $B_2=B_1$. In this case,  parameter $\bm{\gamma}$ will not be updated and the semantic distortion $ -\text{log} \ Q_{\bm{s}_1|\bm{\hat{x}}}(\bm{s}_{1,i}|\bm{\hat{x}}_{ij};\bm{\gamma})$ in \eqref{loss_app} is only regarded as a score to measure how good the recovered image $\hat{\bm{x}}$ contains the semantic information.
	\item \textbf{Step 2}: With fixed parameters $\{\bm{\phi,\psi_1,\psi_2,\theta},\bm{\omega}\}$, train parameter $\bm{\gamma}$.  At this step, we fix the trained parameters $\{\bm{\phi,\psi_1,\psi_2,\theta},\bm{\omega}\}$ and train  parameter $\bm{\gamma}$ for $N_2>0$ epochs to minimize the loss function given in \eqref{loss_app}. Since  parameter $\bm{\gamma}$ is only related to the semantic distortion, the loss function for training the parameter $\bm{\gamma}$ is simplified as 
		\begin{align} \label{loss:final2}
			\tilde{\mathcal{L}}_2=\frac{1}{\hat{B}_2} \sum_{i=1}^{\hat{B}_2} \left\{ \frac{1}{B_n}\sum_{j=1}^{B_n} -\text{log} \ Q_{\bm{s}_1|\bm{\hat{x}}}(\bm{s}_{1,i}|\bm{\hat{x}}_{ij};\bm{\gamma})  \right\},
		\end{align}
		where the batch size $\hat{B}_2$ can be modified to adapt to the different semantic tasks and datasets. 
\end{itemize}

Finally, we repeat the above two steps until the algorithm converges.
In  summary, we present   the iterative training algorithm for the parameters $\{\bm{\phi,\psi_1,\psi_2,\theta},\bm{\gamma},\bm{\omega}\}$ in Algorithm \ref{Training}.

\begin{algorithm}[thb]
		\caption{  Iterative training algorithm for  parameters $\{\bm{\phi,\psi_1,\psi_2,\theta},\bm{\gamma},\bm{\omega}\}$ }
		\label{Training}
		\hrule
		\vspace{0.3cm}
		\begin{algorithmic}[1]
			\Require Training data set $(\mathcal{X},\mathcal{S})$, batch sizes $B_1$, $\hat{B}_2$, and $B_n$, training epochs $N_1$ and $N_2$, weight $\lambda_1$, and maximum  number of iterations $t$.
			\Ensure Trained parameters $\{\bm{\phi}^*,\bm{\psi}^*_1,\bm{\psi}^*_2,\bm{\theta}^*,\bm{\gamma}^*,\bm{\omega}^*\}$.
			
			\State Randomly initial the parameters $\{\bm{\phi,\psi_1,\psi_2,\theta},\bm{\gamma},\bm{\omega}\}$.

		    \Repeat
		    \Repeat
			\State Randomly choose training samples from the data set $(\bm{x}_i,\bm{s}_{1,i})\in(\mathcal{X},\mathcal{S})$ with batch size being $B_1$ and generate the uniform noise samples $o_j$ and $\hat{o}_j$ from PDF $\mathcal{U}(o;0,1)$ with the batch size being $B_n$.
			\State Calculate the loss functions $\tilde{\mathcal{L}}_1$ according to \eqref{loss_app} by letting $B_2=B_1$. 
			\State Fix parameter $\bm{\gamma}$ and calculate the gradients of the loss functions  $\nabla \tilde{\mathcal{L}}_1$ w.r.t.  parameters $\{\bm{\phi,\psi_1,\psi_2,\theta},\bm{\omega}\}$.
			
			\State Update parameters $\{\bm{\phi,\psi_1,\psi_2,\theta},\bm{\omega}\}$ by using the gradient $\nabla \tilde{\mathcal{L}}_1$ via the back propagation algorithm \cite{8054694}.
			\Until The maximum number of training epochs $N_1$ is reached.
			\Repeat 
			\State Randomly choose training samples from the data set $(\bm{x}_i,\bm{s}_{1,i})\in(\mathcal{X},\mathcal{S})$ with batch size being $\hat{B}_2$ and generate the uniform noises $o_j$ and $\hat{o}_j$  from PDF $\mathcal{U}(o;0,1)$ with the batch size being $B_n$.
			\State Calculate the recovered images $\{\hat{\bm{x}}_{ij}\}$ from $\hat{\bm{x}}_{ij}=g_{\bm{\theta}}(f_{\bm{\phi}}(\bm{x}_i)+o_j)$    with the trained parameters $\{\bm{\phi,\psi_1,\psi_2,\theta},\bm{\omega}\}$.
			\State Calculate the loss function $\tilde{\mathcal{L}}_2$ according to \eqref{loss:final2} and its gradient $\nabla \tilde{\mathcal{L}}_2$ w.r.t.  parameter $\bm{\gamma}$.
			\State Update parameter $\bm{\gamma}$ by using the gradient $\nabla \tilde{\mathcal{L}}_2$ via the back propagation algorithm.  
			\Until The maximum number of training epochs $N_2$ is reached.

			\Until The maximum number of iterations $t$ is reached.
            \State  Let $\{\bm{\phi}^*,\bm{\psi}^*_1,\bm{\psi}^*_2,\bm{\theta}^*,\bm{\gamma}^*,\bm{\omega}^*\}=\{\bm{\phi,\psi_1,\psi_2,\theta},\bm{\gamma},\bm{\omega}\}$
		\end{algorithmic}
\end{algorithm}

\section{Simulation Results}

In this section, we present some numerical results to validate the analysis of the proposed FA-based autoencoder scheme for  image transmission and classification.  The experiment setups are given as follows:
\begin{enumerate}
  \item \textbf{Datasets}: To empirically validate our proposed scheme, we conduct the numerical experiments  from a small-scale image dataset with simple labels (CIFAR-10 \cite{CIFAR}) to a large-scale dataset with diverse real-world objects (ImageNet \cite{image}).  Specifically, CIFAR-10 dataset consists of  $50,000$ training images and another $10,000$ validation images with $32 \times 32$ pixels, and its  number of image classes is $10$. ImageNet is a high-resolution image dataset, consisting of $1000$ classes and $1.28$ million training image samples.   During the training and testing, the images of ImageNet  are randomly cropped into $224\times 224$ pixels. Our final results on the ImageNet dataset are evaluated on the $50,000$ validation images.
  \item \textbf{Training details:}  As aforementioned, the network architectures of the functions $f_{\bm{\phi}}$, $g_{\bm{\theta}}$, $h_{\bm{\psi}_1}$, and $h_{\bm{\psi}_2}$ can be used for both the CIFAR-10 and ImageNet datasets. Resnet-$18$ and Resnet-$50$ \cite{he2016deep} are used as the semantic recovery function $Q_{\bm{\gamma}}$ for CIFAR-10 and  ImageNet datasets, respectively. During the model training, we set the mini-batch sizes $B_1=\hat{B}_2=64$ for CIFAR-10 dataset, and mini-batch sizes $B_1=32$ and $\hat{B}_2=256$ for ImageNet dataset. $B_n=1$ for both CIFAR-10 and ImageNet datasets, as the batch sizes $B_1$ and $\hat{B}_2$ are large enough to average out the effects of uniform noise.  PDF $p_{\bm{x}|\bm{\hat{x}}}(\bm{x}|\bm{\hat{x}})=\mathcal{N}(\bm{x}|\bm{\hat{x}},(2\alpha)^{-1}\bm{I})$ is adopted to calculate the image distortion in the loss function \eqref{loss_app}. The parameters $\alpha$ and $\lambda_1$ are adjusted to achieve the different  rate and distortions.    The Adam optimizer \cite{Adam} with a learning rate of $10^{-4}$ is used for updating parameters $\{\bm{\phi,\psi_1,\psi_2,\theta},\bm{\omega}\}$ and the SGD optimizer \cite{8054694} with a learning rate of $10^{-6}$ is used for updating parameter $\{\bm{\gamma}\}$.   Tensorflow 2 \cite{8054694,tensorflow} is employed as the backend.  The proposed iterative algorithm in Algorithm 1 is utilized to train the neural networks. Each iteration consists of $30$ epochs, and  a total of five iterations are performed, resulting in $150$ epochs overall. Our experiments were conducted on a hardware platform equipped with an Intel(R) Xeon(R) Silver 4210R CPU, NVIDIA A100 GPU, and 40GB of RAM. 
  \item \textbf{Comparison schemes}: As comparisons,  the DJSCC methods \cite{kurka2020deepjscc,bourtsoulatze2019deep,dai2022nonlinear} and the classical separate source-channel coding schemes \cite{wallace1992jpeg,sullivan2012overview} are utilized to compress and transmit images. 
  After recovering the images, the trained Resnet-$18$ and Resnet-$50$ are used  to classify the CIFAR-10 and  ImageNet images, respectively.
  In the separate source-channel coding scheme, the classical JPEG method \cite{wallace1992jpeg} and the powerful image codec BPG method\footnote{The performance of the BPG scheme is usually regarded as the benchmark for the image compression problem and is better than some DNN-based schemes, such as  Ball\'{e} \cite{Balle2017}, Luo \cite{luo2018deepsic}, and Theis \cite{theis2017lossy}. } \cite{sullivan2012overview}  are employed as the source coding schemes.  The ideal capacity achieving code \cite{gallager1968information} with bit error probability being zero is considered as the performance bound on the channel coding. In practical implementations, $\frac{2}{3}$ rate $(1944, 1296)$-LDPC code  with $16$-ary quadrature amplitude modulation (16QAM) is used to encode the bits from source coding into the complex symbols.    \end{enumerate}
  \begin{figure}[th] 
	\centering
	\includegraphics[width=3.4in]{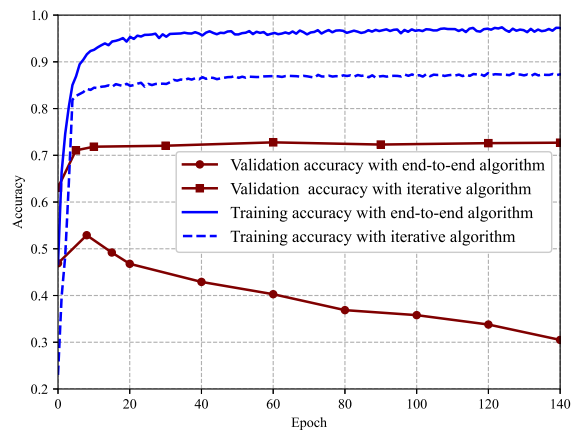}
	\caption{Convergence of the proposed iterative algorithm over ImageNet dataset.}
	\label{simulation_c}
\end{figure}
  \begin{figure*}[t]
\normalsize
\setlength{\abovecaptionskip}{+0.3cm}
\setlength{\belowcaptionskip}{-0.1cm}
	\centering
	\includegraphics[width=6.7in]{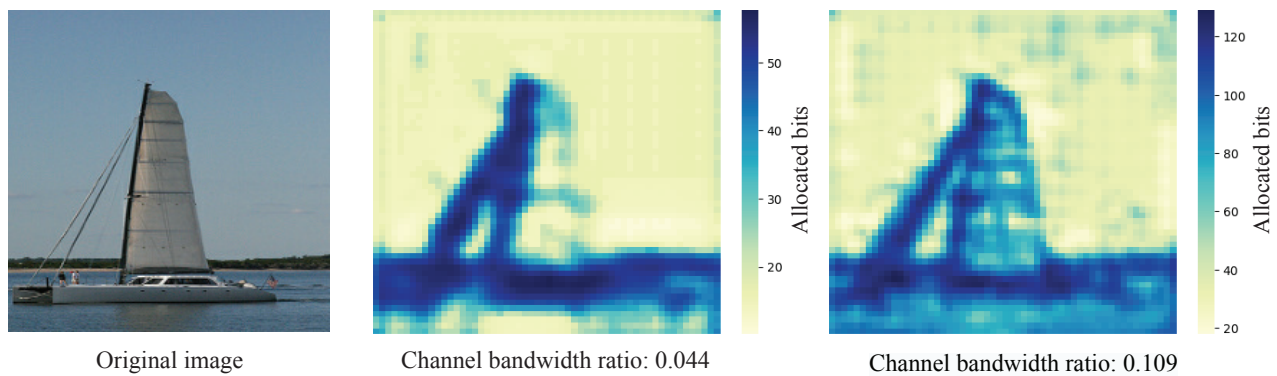}
	\caption{Visualization of  bits allocation for  feature vector $\tilde{\bm{y}}$ over different channel bandwidth ratios. The  original image (left) has $224\times 224$ pixels and both the middle and the right feature images  have $56\times 56$ pixels.  }
	\label{simulation:0}
\end{figure*}

First, we investigate the convergence performance of the proposed iterative training algorithm. To better illustrate the adaptability of the proposed scheme for complex semantic task, we compare it with the traditional end-to-end training algorithm over the ImageNet dataset. The batch size and learning rate for the end-to-end algorithm are set as $32$ and  $10^{-4}$, respectively. Hyperparameters $\lambda_1=1$ and $\alpha=0.3$ are considered. Fig. \ref{simulation_c} plots the classification accuracy as a function of the training epoch. It is observed that the end-to-end training algorithm is incapable of capturing the semantic information of the source data,  resulting in severe overfitting issues.  However, the proposed iterative algorithm addresses this problem and achieves superior accuracy performance on the validation dataset.

Then, we investigate the spatial dependencies among the elements of the extracted feature $\tilde{\bm{y}}$.  Fig. \ref{simulation:0} plots the visualization of the bits allocation for transmitting feature $\tilde{\bm{y}}$ over an additive white Gaussian noise (AWGN) channel with SNR being $10$ dB. Capacity achieving code with rate  $\log(1+10)$ is adopted. The bits at each pixel are calculated as the summation of $-\log P_{\tilde{y}_i|\tilde{z}_i}(\tilde{y}_i|\tilde{z}_i)$ over all the filter elements, i.e., $\sum _{i=1}^{162}-\log P_{\tilde{y}_i|\tilde{z}_i}(\tilde{y}_i|\tilde{z}_i)$, where  PMF $P_{\tilde{y}_i|\tilde{z}_i}(\tilde{y}_i|\tilde{z}_i)$ is calculated from \eqref{den:y}. From Fig. \ref{simulation:0}, it is observed that 
the pixels of the same object, e.g., boat, sky, and water, have similar bits, which reveals that these elements of  feature $\tilde{\bm{y}}$ are spatially correlated. In addition, we observe that when the channel bandwidth ratio increases, more bits are utilized to characterize the details of the objects, and thus the recovered image is more close to the original image.
\begin{figure}[htbp]
\centering

\subfigure[PSNR]{
\begin{minipage}[t]{1\linewidth}
	\centering
	\includegraphics[width=3.5in]{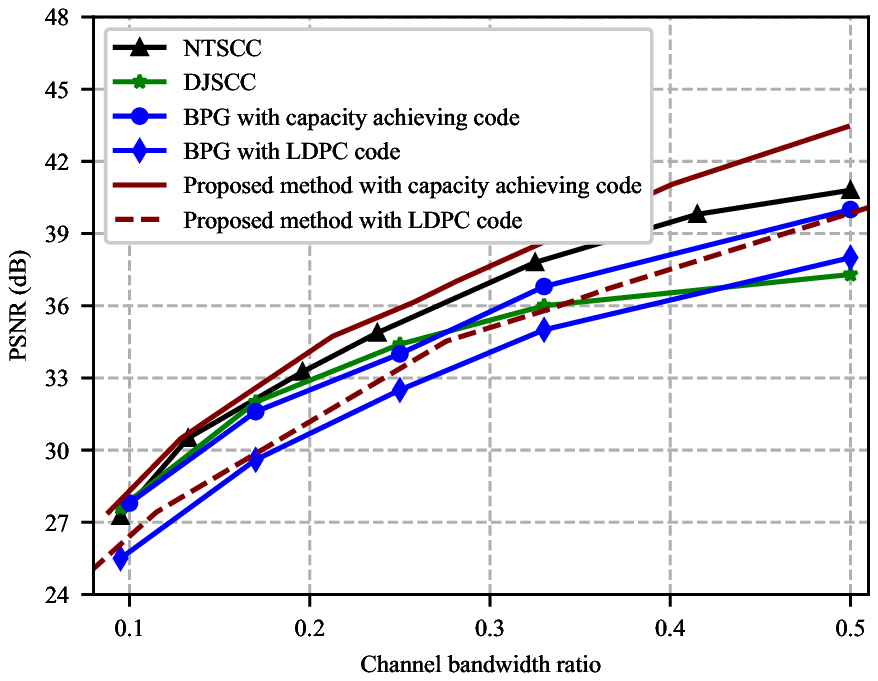}
%\caption{fig1}
\end{minipage}%
}%

\subfigure[MS-SSIM]{
\begin{minipage}[t]{1\linewidth}
	\centering
	\includegraphics[width=3.5in]{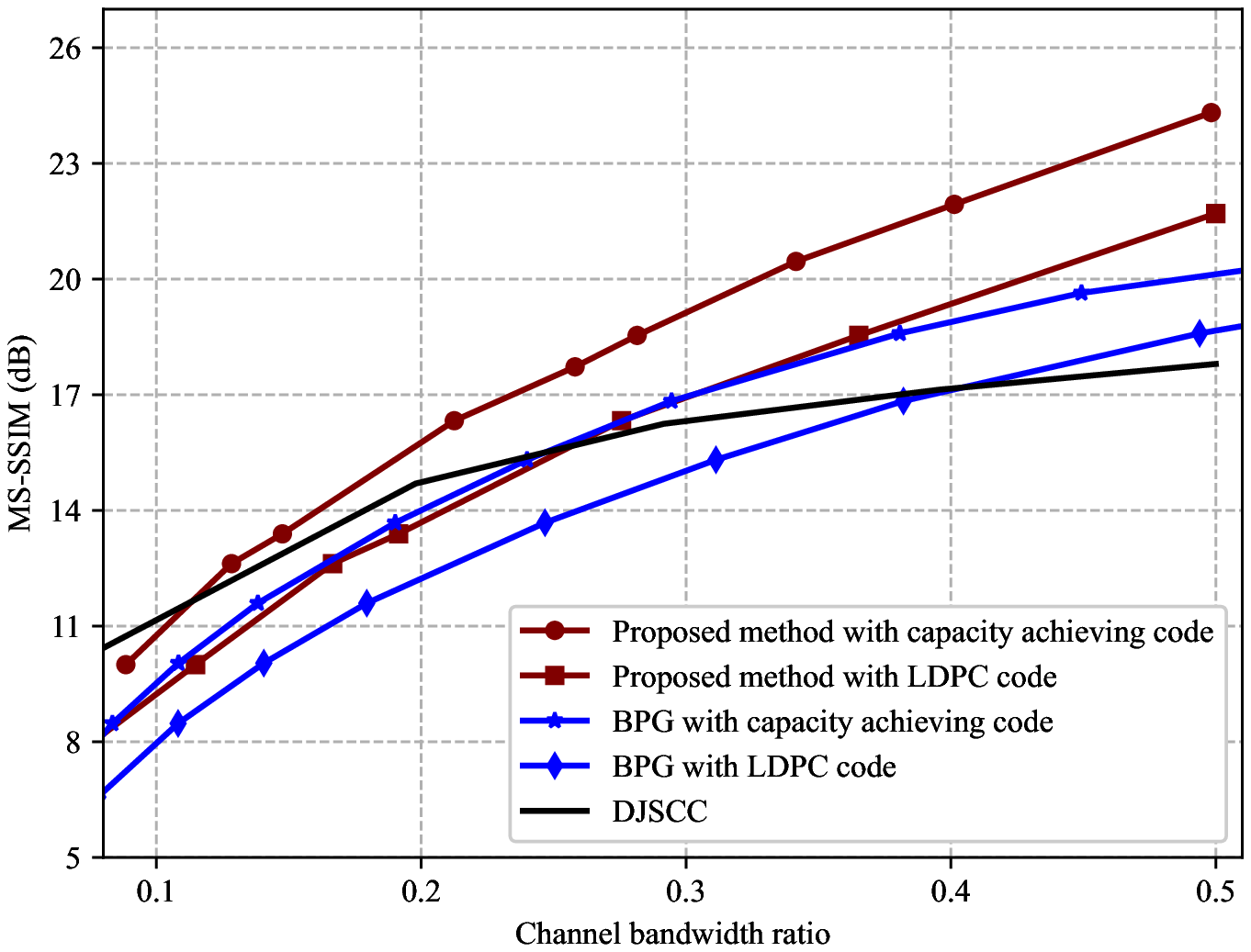}
%\caption{fig2}
\end{minipage}%
}%
\centering
\caption{Performance comparisons of the proposed scheme with the state-of-art ones over CIFAR-10 dataset with  SNR being $10$ dB.}
\label{simulation:1}
\end{figure}

Next, we compare the performance of our proposed scheme with the DJSCC and BPG schemes over CIFAR-10 dataset. Here, we do not consider JPEG scheme since its image recovery and classification performance are far inferior to the other schemes over CIFAR-10 dataset.  The widely used pixel-wise metric, i.e., the peak signal-to-noise ratio (PSNR),  and the perceptual metric, i.e., the multi-scale structural similarity index (MS-SSIM), are utilized to measure the performance of  image recovery.  Fig. \ref{simulation:1}(a) plots PSNR as a function of channel bandwidth ratio over an AWGN channel with SNR being $10$ dB. It is observed that our proposed scheme with capacity achieving code outperforms the nonlinear transform source-channel coding (NTSCC) \cite{dai2022nonlinear}, the DJSCC method \cite{bourtsoulatze2019deep}, and the BPG scheme. For example, when the channel bandwidth ratio is $0.4$, the proposed scheme outperforms the NTSCC and the BPG scheme with capacity achieving code by around $1.5$ dB and $2.7$ dB, respectively. In addition, it is observed that adopting  LDPC as the channel code  degrades the performance of both the FA-based  autoencoder and BPG schemes since the rate of channel coding is decreased. However, the proposed scheme still outperforms the BPG scheme and surpasses the DJSCC scheme when the channel bandwidth ratio is large. Fig. \ref{simulation:1}(b) plots  MS-SSIM as a function of  channel bandwidth ratio. It is observed that the proposed scheme has a superior performance over the state-of-art schemes. For instance, when the channel bandwidth ratio is $0.3$, the proposed scheme with capacity achieving code outperforms  BPG scheme with capacity achieving code and  DJSCC  scheme by around $2.5$ dB and $2.7$ dB, respectively.

  \begin{figure}[th] 
	\centering
	\includegraphics[width=3.4in]{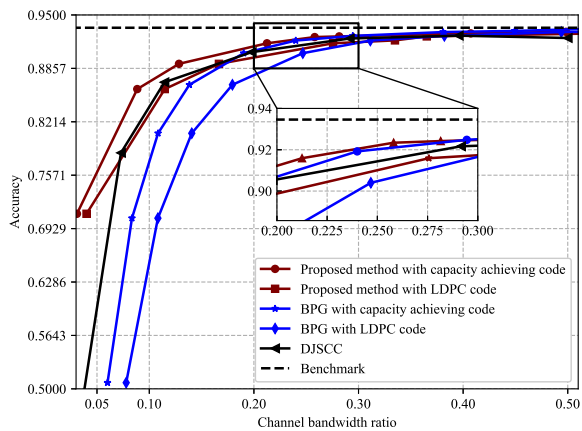}
	\caption{Classification accuracy versus channel bandwidth ratio over  CIFAR-10 dataset with SNR being $10$ dB.}
	\label{simulation:3}
\end{figure}

Fig. \ref{simulation:3} plots the classification accuracy as a function of  channel bandwidth ratio over  CIFAR-10 dataset with  SNR being $10$ dB. The benchmark results are obtained by employing a pretrained Resnet-$18$ on the original images.  It is observed that 
the proposed scheme with capacity achieving code has a higher classification accuracy than the DJSCC  and  BPG schemes. For instance, when the channel bandwidth ratio is $0.05$, the classification accuracy of the proposed scheme with capacity achieving code is $17.5\%$ higher than that of the DJSCC scheme.  Furthermore, when the channel bandwidth decreases, the performance gain of the proposed scheme becomes larger, since other schemes only focus on the image recovery and do not consider the semantic distortion. As the channel bandwidth ratio increases, the accuracies of all the schemes approach to that of the benchmark scheme, since the recovered images are close to the original ones.

\begin{figure}[htbp]
\centering

\subfigure[PSNR]{
\begin{minipage}[t]{1\linewidth}
	\centering
	\includegraphics[width=3.3in]{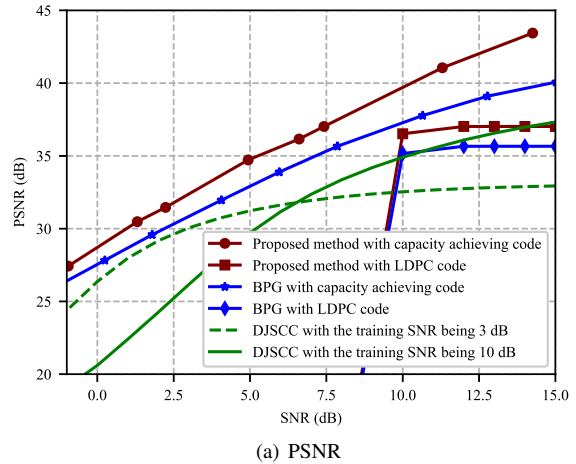}
%\caption{fig1}
\end{minipage}%
}%

\subfigure[Classification accuracy]{
\begin{minipage}[t]{1\linewidth}
	\centering
	\includegraphics[width=3.3in]{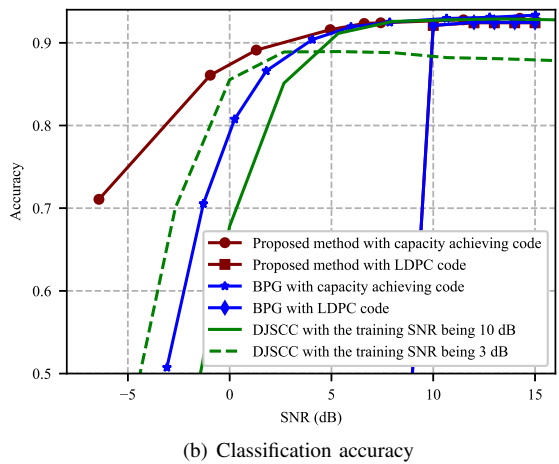}
%\caption{fig2}
\end{minipage}%
}%
\centering
\caption{Performance comparisons of the proposed scheme with the state-of-art ones over CIFAR-10 dataset, where the channel bandwidth ratio is fixed as $0.36$. }
\label{simulation:4}
\end{figure}

   \begin{figure}[th] 
	\centering
	\includegraphics[width=3.5in]{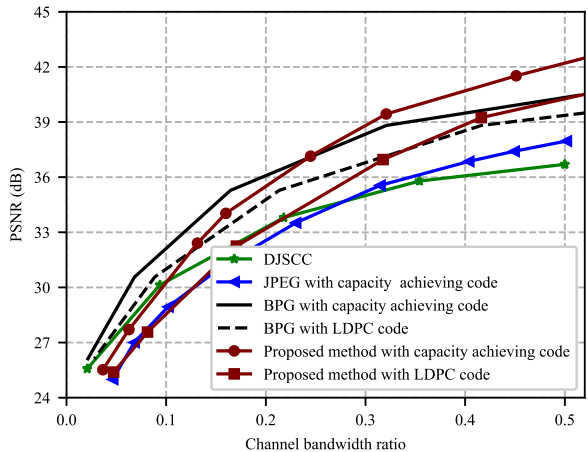}
	\caption{PSNR performance versus channel bandwidth ratio over  ImageNet dataset with SNR being $10$ dB.}
	\label{Image:1}
\end{figure}

\begin{figure}[htbp]
\centering

\subfigure[Top-1 accuracy]{
\begin{minipage}[t]{1\linewidth}
	\centering
	\includegraphics[width=3.2 in]{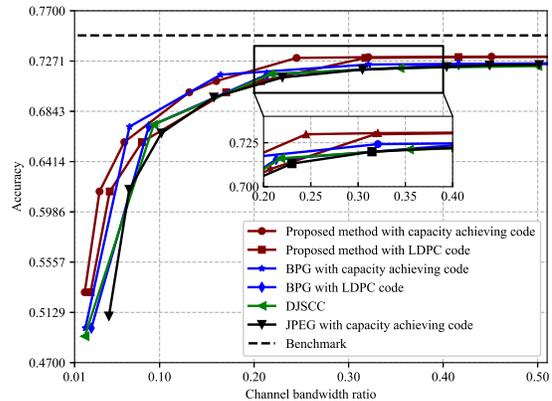}
%\caption{fig1}
\end{minipage}%
}%

\subfigure[Top-5 accuracy]{
\begin{minipage}[t]{1\linewidth}
	\centering
	\includegraphics[width=3.2 in]{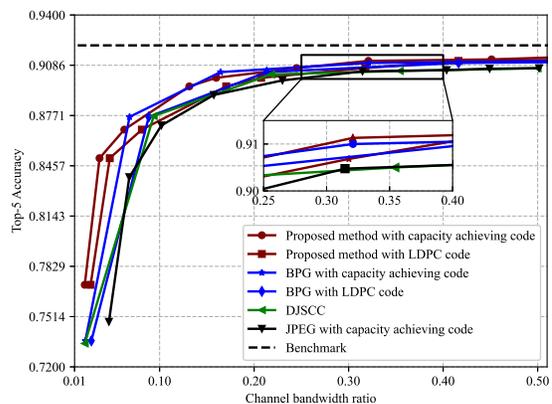}
%\caption{fig2}
\end{minipage}%
}%
\centering
\caption{Classification accuracy versus  channel bandwidth ratio over  ImageNet dataset with SNR being $10$ dB.}
\label{Image:3}
\end{figure}

Then, we investigate the performance of the proposed scheme with  different SNRs. To fairly compare the performances of all the schemes, the channel bandwidth ratio is fixed as $0.36$.  Fig. \ref{simulation:4}(a) plots  PSNR as a function of SNR over an AWGN channel. It is observed that our proposed scheme with capacity achieving code has a superior performance over all the state-of-art schemes. For example, when the SNR is $5$ dB, the proposed scheme with capacity achieving code outperforms the BPG scheme with capacity achieving code and DJSCC scheme with the training SNR being $10$ dB by around $1.6$ dB and $6.6$ dB, respectively. The reason why the DSSCC scheme is superior to the DJSCC scheme is that  the DJSCC scheme suffers from the performance loss caused by SNR mismatch. However, adopting the LDPC code will significantly degrade the performance of the proposed DSSCC scheme when the SNR is low, and this phenomenon is known as the \emph{cliff effect}. The reason to explain this phenomenon is that when the SNR decreases, bit errors increase and propagate among the coded bits, which severely corrupts the feature map. To mitigate the cliff effect, future research will explore using advanced channel coding or resource allocation to transmit features with different degrees of importance.   However, when the error probability is acceptable, the proposed scheme with the LDPC code still outperforms the DJSCC method. For instance, when the SNR is $12$ dB, the proposed scheme with LDPC code outperforms the DJSCC scheme with the training SNR being $3$ dB by around $4.3$ dB.  Fig. \ref{simulation:4}(b) plots the classification accuracy as a function of  SNR over an AWGN channel with channel bandwidth ratio being $0.36$. It is observed that  the proposed scheme with capacity achieving code has a superior performance over all the schemes, especially for the low SNR regimes. For instance, when the SNR is $-5$ dB, the accuracy of the proposed scheme with capacity achieving code is $25\%$ higher than that of the DJSCC scheme with the training SNR being $3$ dB. In addition, when the SNR is high, the DJSCC with lower training SNR has lower accuracy than the proposed scheme with LDPC code. The performance loss is caused by the mismatch between the training  and testing SNRs.

\begin{figure*}[t]
\normalsize
\setlength{\abovecaptionskip}{+0.3cm}
\setlength{\belowcaptionskip}{-0.1cm}
	\centering
	\includegraphics[width=5.0 in]{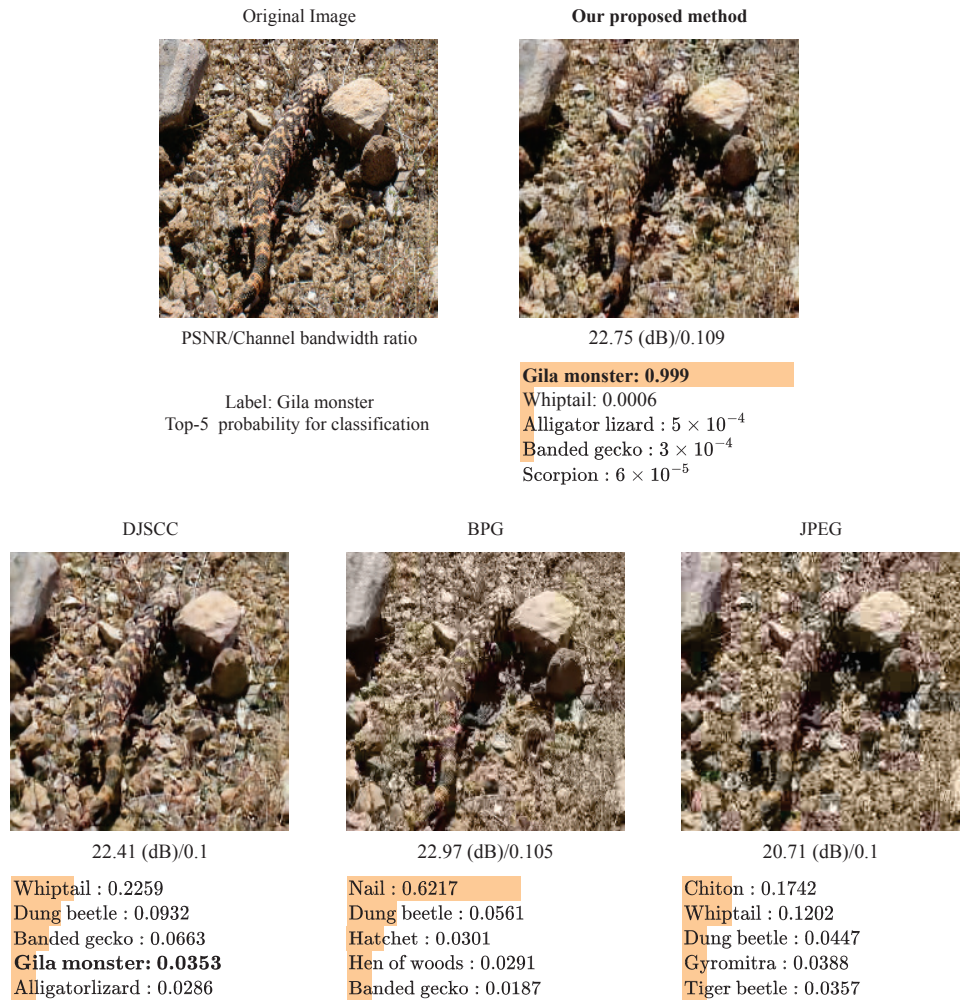}
	\caption{ An example for the performance comparison over ImageNet dataset with  SNR being $10$ dB. }
	\label{Image:2}
\end{figure*}

Finally, we investigate the performance of the proposed scheme over ImageNet dataset. Fig. \ref{Image:1} plots  PSNR as a function of the channel bandwidth ratio over an AWGN channel with  SNR being $10$ dB. It is observed that our proposed scheme still has a superior performance over the DJSCC  and  JPEG schemes. For instance, when the channel bandwidth ratio is $0.3$, the proposed schemes with capacity achieving code and LDPC code outperform the DJSCC scheme by around $3.7$ dB and $1.2$ dB, respectively.  However,  the PSNR performance of the proposed method is lower than that of BPG under low channel bandwidth ratio conditions. The reason to explain this phenomenon is that the adopted forward adaptation scheme requires the additional channel bandwidth to transmit the spatial distribution of the image, which degrades the PSNR performance when the bandwidth ratio is low. To tackle this problem, more advanced architectures of the neural networks can be explored to improve the  efficiency of compressing the spatial distribution. 
 Fig. \ref{Image:3} plots the classification accuracy as a function of  channel bandwidth ratio over ImageNet dataset. Here, top-$5$ accuracy is also adopted as the metric to measure the classification performance, which considers a classification correct if  any predicted labels of the top-$5$ highest probability  match with the target label. As shown in Fig. \ref{Image:3}(a) and Fig. \ref{Image:3}(b), it is observed that when the channel bandwidth ratio is small, the proposed scheme has higher classification accuracy than the DJSCC and JPEG schemes under both the capacity achieving code and the LDPC code. For instance,  when the channel bandwidth ratio is $0.02$, the top-5 accuracy of the  proposed scheme with LDPC code is $1.8\%$ higher than the DJSCC scheme. It is worth noticing that the BPG scheme outperforms the proposed scheme in the medium channel bandwidth regime, since the BPG scheme has better image recovery performance in these regimes as shown in Fig. \ref{Image:1}. However, with the decreasing of the channel bandwidth ratio, the proposed scheme that considers the semantic distortion still have a superior performance than the BPG scheme.  As shown in Fig. \ref{Image:2}, an image labeled as ``Gila monster" is presented and the performance of our proposed method, BPG,  JPEG, and  DJSCC schemes  are compared.  Here, capacity achieving code is used as the channel code for the separate source-channel coding schemes. It is observed that with similar channel bandwidth ratio, our proposed scheme has a higher PSNR performance than the DJSCC  and  JPEG schemes. In our proposed scheme, the probability of detecting the ``Gila monster" label approaches $1$, which is much larger than that of the other schemes. 

%Finally, we compare the proposed scheme with some semantic-aware DSSCC schemes in the deep learning community. We consider the AWGN channel with the SNR being $10$ dB and capacity achieving code with rate  $\log(1+10)$ is adopted. As shown in Table \ref{Comparison one}, it is observed that with less channel bandwidth ratio 

%\begin{table} [!htbp]
%\newcommand{\tabincell}[2]{\begin{tabular}{@{}#1@{}}#2\end{tabular}}
% \centering
% \caption{Performance comparisons of the semantic-aware DSSCC schemes  over ImageNet dataset.}  
% \begin{spacing}{1.1}
%  \begin{tabu}{p{2.0cm}<{\centering}|p{2.5cm}<{\centering}|p{1.2cm}<{\centering}|p{1.2cm}<{\centering}}
%   \tabucline[1.9pt]{-}
%   \textbf{Methods} & \textbf{Bandwidth ratio}  & \textbf{PSNR} & \textbf{Accuracy} \\
%   \tabucline[1.5pt]{-}
%   \hline  
%   DeepSIC \cite{luo2018deepsic} & 0.08 & 30.6 & 52.2\% \\
%   \hline    
%   GMM-E \cite{kawawa2022recognition} & 0.08 & 20.7 & 65\% \\
%   \hline
%   Our method & \textbf{0.0626} & \textbf{27.7} & \textbf{65.8}\%\\ 
%   \hline 
%   \tabucline[1.5pt]{-}
%  \end{tabu}  
% \end{spacing}  
% \label{Comparison one}
%\end{table}

\setcounter{TempEqCnt1}{\value{equation}}
\setcounter{TempEqCnt}{\value{equation}}
\begin{figure*}[b]
		  \hrulefill
		\vspace{-5pt}
		\setcounter{equation}{32}
		\begin{align}
	\label{proof:eq1}	\mathcal{L}_{\bm{\phi,\theta}} &= \mathbb{E}_{\bm{x}} \left\{ \mathcal{K} \left(q_{\tilde{\bm{y}}|\bm{x}}||  p_{\tilde{\bm{y}}|\bm{x}} \right)\right\}+ \sum_{i=1}^{f}\lambda_i\mathbb{E}_{\bm{s}_i}\left\{ \mathcal{K} \left(q_{\tilde{\bm{y}}|\bm{s}_i}||  p_{\tilde{\bm{y}}|\bm{s}_i }    \right) \right\},\\
	\label{proof:eq2}	&=\mathbb{E}_{\bm{x}} \mathbb{E}_{\tilde{\bm{y}}\sim q_{\tilde{\bm{y}}|\bm{x}}}\left\{\log \frac{q_{\tilde{\bm{y}}|\bm{x}}}{p_{\tilde{\bm{y}}|\bm{x}}} \right\}+\sum_{i=1}^{f}\lambda_i \mathbb{E}_{\bm{s}_i} \mathbb{E}_{\tilde{\bm{y}}\sim q_{\tilde{\bm{y}}|\bm{s}_i}}\left\{\log \frac{q_{\tilde{\bm{y}}|\bm{s}_i}}{p_{\tilde{\bm{y}}|\bm{s}_i } } \right\},\\
	\label{proof:eq3}	&= \mathbb{E}_{\bm{x}} \mathbb{E}_{\tilde{\bm{y}}\sim q_{\tilde{\bm{y}}|\bm{x}}}\left\{\log q_{\tilde{\bm{y}}|\bm{x}}-\log p_{\bm{x}|\tilde{\bm{y}}} -\log p_{\tilde{\bm{y}}}(\tilde{\bm{y}})+\log p_{\bm{x}}(\bm{x})\right\}\nonumber\\
		&\quad + \sum_{i=1}^{f}\lambda_i \mathbb{E}_{\bm{s}_i} \mathbb{E}_{\tilde{\bm{y}}\sim q_{\tilde{\bm{y}}|\bm{s}_i}}
		\left\{\log q_{\tilde{\bm{y}}|\bm{s}_i}-\log p_{\bm{s}_i|\tilde{\bm{y}}} -\log p_{\tilde{\bm{y}}}(\tilde{\bm{y}})+\log p_{\bm{s}_i}(\bm{s}_i)\right\},\\
	\label{proof:eq4}	&= \mathbb{E}_{\bm{x}} \mathbb{E}_{\tilde{\bm{y}}\sim q_{\tilde{\bm{y}}|\bm{x}}}\left\{ -\log p_{\bm{x}|\bm{\hat{x}}}(\bm{x}|\bm{\hat{x}})\right\}
	+ \mathbb{E}_{\bm{x}} \mathbb{E}_{\tilde{\bm{y}}\sim q_{\tilde{\bm{y}}|\bm{x}}}\left\{ -\log p_{\tilde{\bm{y}}}(\tilde{\bm{y}})\right\}\nonumber\\
	&\quad + \sum_{i=1}^{f}\lambda_i \mathbb{E}_{\bm{s}_i} \mathbb{E}_{\tilde{\bm{y}}\sim q_{\tilde{\bm{y}}|\bm{s}_i}}
		\left\{ -\log p_{\bm{s}_i|\bm{\hat{x}}}(\bm{s}_i|\bm{\hat{x}})\right\}+\lambda_i \mathbb{E}_{\bm{s}_i} \mathbb{E}_{\tilde{\bm{y}}\sim q_{\tilde{\bm{y}}|\bm{s}_i}}
		\left\{ -\log p_{\tilde{\bm{y}}}(\tilde{\bm{y}})\right\}\nonumber\\
		&\quad +\underbrace{ \mathbb{E}_{\bm{x}} \mathbb{E}_{\tilde{\bm{y}}\sim q_{\tilde{\bm{y}}|\bm{x}}}\left\{\log q_{\tilde{\bm{y}}|\bm{x}}\right\}}_{\text{\ding{172}}} + \sum_{i=1}^{f}\lambda_i \underbrace{ \mathbb{E}_{\bm{s}_i} \mathbb{E}_{\tilde{\bm{y}}\sim q_{\tilde{\bm{y}}|\bm{s}_i}}\left\{\log q_{\tilde{\bm{y}}|\bm{s}_i}\right\}}_{\text{\ding{173}}}\nonumber \\
	&\quad +\underbrace{\mathbb{E}_{\bm{x}}\{\log p_{\bm{x}}(\bm{x})\}}_{\text{\ding{174}}}+\sum_{i=1}^{f}\lambda_i\underbrace{\mathbb{E}_{\bm{s}_i}\{\log p_{\bm{s}_i}(\bm{s}_i)\}}_{\text{\ding{175}}},\\ 
		& \leq \mathbb{E}_{\bm{x}} \mathbb{E}_{\tilde{\bm{y}}\sim q_{\tilde{\bm{y}}|\bm{x}}} \left\{-\text{log} \ p_{\bm{x}|\bm{\hat{x}}}(\bm{x}|\bm{\hat{x}})  \right\}+\sum_{i=1}^{f}\lambda_i \mathbb{E}_{\bm{s}_i} \mathbb{E}_{\tilde{\bm{y}}\sim q_{\tilde{\bm{y}}|\bm{s}_i}} \left\{-\text{log} \ p_{\bm{s}_i|\bm{\hat{x}}}(\bm{s}_i|\bm{\hat{x}})  \right\}\nonumber \\
	\label{proof:eq5} &\quad + (1+\lambda)\mathbb{E}_{\bm{x}} \mathbb{E}_{\tilde{\bm{y}}\sim q_{\tilde{\bm{y}}|\bm{x}}}\left\{-\log p(\tilde{\bm{y}})\right\} +\mathcal{T},
\end{align}
\setcounter{equation}{42}
	\begin{align}
	\label{proof:equ3_4_0}		\mathbb{E}_{\bm{x}} \mathbb{E}_{\tilde{\bm{y}}\sim q_{\tilde{\bm{y}}|\bm{x}}}\left\{-\log p(\tilde{\bm{y}})\right\}&=\int_{\bm{x}}\int_{\tilde{\bm{y}}}-p_{\bm{x}}(\bm{x})q_{\tilde{\bm{y}}|\bm{x}}(\tilde{\bm{y}}|\bm{x};\bm{\phi})\log p_{\tilde{\bm{y}}}(\tilde{\bm{y}})d\tilde{\bm{y}}d\bm{x},\\
	\label{proof:equ3_4_1}		&=\int_{\bm{x}}\int_{\bm{s}_i}\int_{\tilde{\bm{y}}}-p_{\bm{x},\bm{s}_i}(\bm{x},\bm{s}_i)q_{\tilde{\bm{y}}|\bm{x}}(\tilde{\bm{y}}|\bm{x};\bm{\phi})\log p_{\tilde{\bm{y}}}(\tilde{\bm{y}})d\tilde{\bm{y}}d{\bm{s}_i}d\bm{x}.
		\end{align}
		\begin{align}
			\mathbb{E}_{\bm{s}_i} \mathbb{E}_{\tilde{\bm{y}}\sim q_{\tilde{\bm{y}}|\bm{s}_i}}\left\{-\log p(\tilde{\bm{y}})\right\}&=\int_{\bm{s}_i}\int_{\tilde{\bm{y}}}-p_{\bm{s}_i}(\bm{s}_i)q_{\tilde{\bm{y}}|\bm{s}_i}(\tilde{\bm{y}}|\bm{s}_i;\bm{\phi})\log p_{\tilde{\bm{y}}}(\tilde{\bm{y}})d\tilde{\bm{y}}d{\bm{s}_i},\\
	\label{proof:equ3_2}		&=\int_{\bm{s}_i} \int_{\tilde{\bm{y}}}-p_{\bm{s}_i}(\bm{s}_i)\left\{\int_{\bm{x}} q_{\bm{\tilde{y}}|\bm{x}} \cdot p_{\bm{x}|\bm{s}_i}(\bm{x}|\bm{s}_i)d\bm{x}\right\}\log p_{\tilde{\bm{y}}}(\tilde{\bm{y}})d\tilde{\bm{y}} d{\bm{s}_i},\\
		\label{proof:equ3_4}	&= \int_{\bm{x}}\int_{\bm{s}_i}\int_{\tilde{\bm{y}}}-p_{\bm{x},\bm{s}_i}(\bm{x},\bm{s}_i)q_{\tilde{\bm{y}}|\bm{x}}(\tilde{\bm{y}}|\bm{x};\bm{\phi})\log p_{\tilde{\bm{y}}}(\tilde{\bm{y}})d\tilde{\bm{y}}d{\bm{s}_i}d\bm{x}.
		\end{align}
		\vspace{-5pt}
		\setcounter{TempEqCnt}{\value{equation}}
\end{figure*}

\section{Concluding Remarks}
This paper proposed an efficient DSSCC-based autoencoder approach for the JTD-SC problem, by exploring the extended rate-distortion theory with semantic distortion.  First, we derived the  rate-distortion optimization function with semantic distortion for general source data and semantic tasks. By taking image transmission and classification as example, we proposed an FA-based autoencoder scheme, which 
involves the adaptive density module to learn the PDF  of the image features. Finally, in order to tackle the overfitting problem, we proposed an iterative training algorithm to iteratively train the neural networks for recovering data and semantic information.  Simulation results revealed that the proposed scheme with capacity achieving code has superior performances, in terms of PSNR and classification accuracy, over the  BPG and DJSCC schemes combined with the trained classification DNNs. However, the proposed scheme with LDPC code suffered from the cliff effect, which degrades the  performance of both the data and label recovery when the SNR is low. Notably, the joint optimization of the autoencoder and channel coding can reduce the effect of the performance degradation, which will be left for the future studies.

\appendices
\section{Proof of Proposition \ref{pro:1}}
%\setcounter{TempEqCnt1}{\value{equation}}
%\setcounter{TempEqCnt}{\value{equation}}
%	\begin{figure*}[t]
%		\vspace{-5pt}
%		\setcounter{equation}{30}
%
%		\vspace{-5pt}
%		\setcounter{TempEqCnt}{\value{equation}}
%		\hrulefill
%\end{figure*}
\setcounter{equation}{37}
From equation \eqref{equ:1}, we have \eqref{proof:eq1}-\eqref{proof:eq5}, where $\lambda=\sum_{i=1}^{f}\lambda_i$ and $\mathcal{T}=\mathbb{E}_{\bm{x}}\{\log p_{\bm{x}}(\bm{x})\}+\sum_{i=1}^{f}\lambda_i\mathbb{E}_{\bm{s}_i}\{\log p_{\bm{s}_i}(\bm{s}_i)\}$ is constant, \eqref{proof:eq2} comes from the definition of KL divergence \cite{gallager1968information}, \eqref{proof:eq3} comes from the Bayesian expression $p_{\bm{\tilde{y}}|\bm{x}}(\bm{\tilde{y}}|\bm{x})=p_{\bm{x}|\bm{\tilde{y}}}(\bm{x}|\bm{\tilde{y}})p_{\bm{\tilde{y}}}(\bm{\tilde{y}})/p_{\bm{x}}(\bm{x})$, and \eqref{proof:eq4} comes from \eqref{equ:pos1} and \eqref{equ:pos2}. In the followings, we prove that \eqref{proof:eq5} holds. First,  we show that term \ding{173} $\leq$ term \ding{172} $=0$, which results in the inequality in \eqref{proof:eq5}. Then we show that terms \ding{174} and \ding{175} are constant, and 
 $\mathbb{E}_{\bm{x}} \mathbb{E}_{\tilde{\bm{y}}\sim q_{\tilde{\bm{y}}|\bm{x}}}\left\{-\log p(\tilde{\bm{y}})\right\}=\mathbb{E}_{\bm{s}_i} \mathbb{E}_{\tilde{\bm{y}}\sim q_{\tilde{\bm{y}}|\bm{s}_i}}\left\{-\log p(\tilde{\bm{y}})\right\}$, which completes the proof.
\begin{itemize}
\item Term \ding{173} $\leq $ term \ding{172}: Terms \ding{172} and  \ding{173} are the negative conditional entropy of the probabilities $q_{\bm{\tilde{y}}|\bm{x}}$ and $q_{\bm{\tilde{y}}|\bm{s}}$, respectively, and they are denoted as
\begin{align}
	\label{H:1}\mathbb{E}_{\bm{s}} \mathbb{E}_{\tilde{\bm{y}}\sim q_{\tilde{\bm{y}}|\bm{s}}}\left\{\log q_{\tilde{\bm{y}}|\bm{s}}\right\} &= -H_{q}(\tilde{\bm{y}}|\bm{s}_i),\\
	 \label{H:2}\mathbb{E}_{\bm{x}} \mathbb{E}_{\tilde{\bm{y}}\sim q_{\tilde{\bm{y}}|\bm{x}}}\left\{\log q_{\tilde{\bm{y}}|\bm{x}}\right\} &= -H_{q}(\tilde{\bm{y}}|\bm{x}),
\end{align}
where $H_q(\bm{X}|\bm{Y})$ indicates  the conditional entropy of the random variable $\bm{X}$ given $\bm{Y}$ with the conditional  PDF being $q_{\bm{X}|\bm{Y}}(\bm{X}|\bm{Y})$. According to the Markov chain in Fig. \ref{Appendix:2}(a) and data processing inequality \cite{gallager1968information}, we have 
\begin{align}
	H(\tilde{\bm{y}})-H_{q}(\tilde{\bm{y}}|\bm{s}_i)\leq H(\tilde{\bm{y}})-H_{q}(\tilde{\bm{y}}|\bm{x}),
\end{align}
which implies $-H_{q}(\tilde{\bm{y}}|\bm{s}_i) \leq -H_{q}(\tilde{\bm{y}}|\bm{x})$. Hence, we have Term \ding{173} $\leq $ term \ding{172} from \eqref{H:1} and \eqref{H:2}.

	\item Term \ding{172}: Under the error-free channel transmission, the approximated probability $q_{\bm{\tilde{y}}|\bm{x}}$ follows
	\begin{align} \label{proof:equ1_1}
	q_{\bm{\tilde{y}}|\bm{x}}(\bm{\tilde{y}}|\bm{x};\bm{\phi})&=	\prod \limits_{i=1}^{K}\mathcal{U}(\tilde{y}_i;y_i,1), \ \text{with}\ \bm{y}=f_{\bm{\phi}}(\bm{x}),\\
	&=\left\{\begin{array}{ll}
	1, & y_i-0.5\leq \tilde{y}_i \leq y_i+0.5, \forall i,\\
	0, & \text{otherwise}.
\end{array}\right.
	\end{align}
	Hence, term \ding{172} is the sum of the entropies of the uniform distribution which evaluates to zero.
	
\item $\mathbb{E}_{\bm{x}} \mathbb{E}_{\tilde{\bm{y}}\sim q_{\tilde{\bm{y}}|\bm{x}}}\left\{-\log p_{\tilde{\bm{y}}}(\tilde{\bm{y}})\right\}=\mathbb{E}_{\bm{s}_i} \mathbb{E}_{\tilde{\bm{y}}\sim q_{\tilde{\bm{y}}|\bm{s}_i}}\left\{-\log p(\tilde{\bm{y}})\right\}$: From \eqref{proof:eq5}, we have \eqref{proof:equ3_4_0}-\eqref{proof:equ3_4}. \eqref{proof:equ3_2} holds due to the fact that 
\setcounter{equation}{47}
\begin{align}
			q_{\bm{\tilde{y}}|\bm{s}_i}(\bm{\tilde{y}}|\bm{s}_i;\bm{\phi})&=\int_{\bm{x}} q_{\bm{\tilde{y}},\bm{x}|\bm{s}_i}(\bm{\tilde{y}},\bm{x}|\bm{s}_i;\bm{\phi})d\bm{x},\\
		&=\int_{\bm{x}} q_{\bm{\tilde{y}}|\bm{x},\bm{s}_i}(\bm{\tilde{y}}|\bm{x},\bm{s}_i;\bm{\phi})p_{\bm{x}|\bm{s}_i}(\bm{x}|\bm{s}_i)d\bm{x},\\
	\label{proof:equ2_2}	&=\int_{\bm{x}} q_{\bm{\tilde{y}}|\bm{x}}(\bm{\tilde{y}}|\bm{x};\bm{\phi})p_{\bm{x}|\bm{s}_i}(\bm{x}|\bm{s}_i)d\bm{x},
		\end{align}
		where \eqref{proof:equ2_2} comes from  the fact that $\bm{\tilde{y}}$ is only determined by $\bm{x}$ according to the Markov chain in Fig. \ref{Appendix:2}. In addition, \eqref{proof:equ3_4} holds by changing the integral order. According to \eqref{proof:equ3_4_1} and \eqref{proof:equ3_4}, we have $\mathbb{E}_{\bm{x}} \mathbb{E}_{\tilde{\bm{y}}\sim q_{\tilde{\bm{y}}|\bm{x}}}\left\{-\log p_{\tilde{\bm{y}}}(\tilde{\bm{y}})\right\}=\mathbb{E}_{\bm{s}_i} \mathbb{E}_{\tilde{\bm{y}}\sim q_{\tilde{\bm{y}}|\bm{s}_i}}\left\{-\log p(\tilde{\bm{y}})\right\}$.
	\end{itemize}
Hence, Proposition \ref{pro:1} is proved.

\IEEEpeerreviewmaketitle
\bibliographystyle{IEEEtran}
\bibliography{semantic}
\begin{IEEEbiography}[{\includegraphics[width=1in,height=1.25in,clip,keepaspectratio]{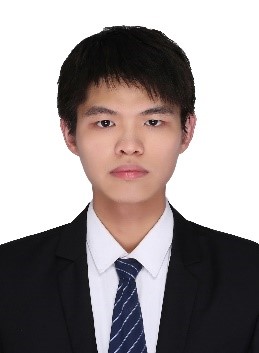}}]{Jianhao Huang} received his B.S. degree in electrical engineering from Harbin Engineering University, China, in 2017.  He is currently pursuing the Ph.D. degree in the School of Science and Engineering, The Chinese University of Hong Kong, Shenzhen. He was a TPC member for IEEE GLOBECOM 2019-2022. He was a reviewer for IEEE Wireless Communications Letters. He received the best paper award from IEEE GLOBECOM 2020. His current research interests include compress sensing in communication systems, semantic communications, and deep learning.\end{IEEEbiography}

\begin{IEEEbiography}[{\includegraphics[width=1in,height=1.25in,clip,keepaspectratio]{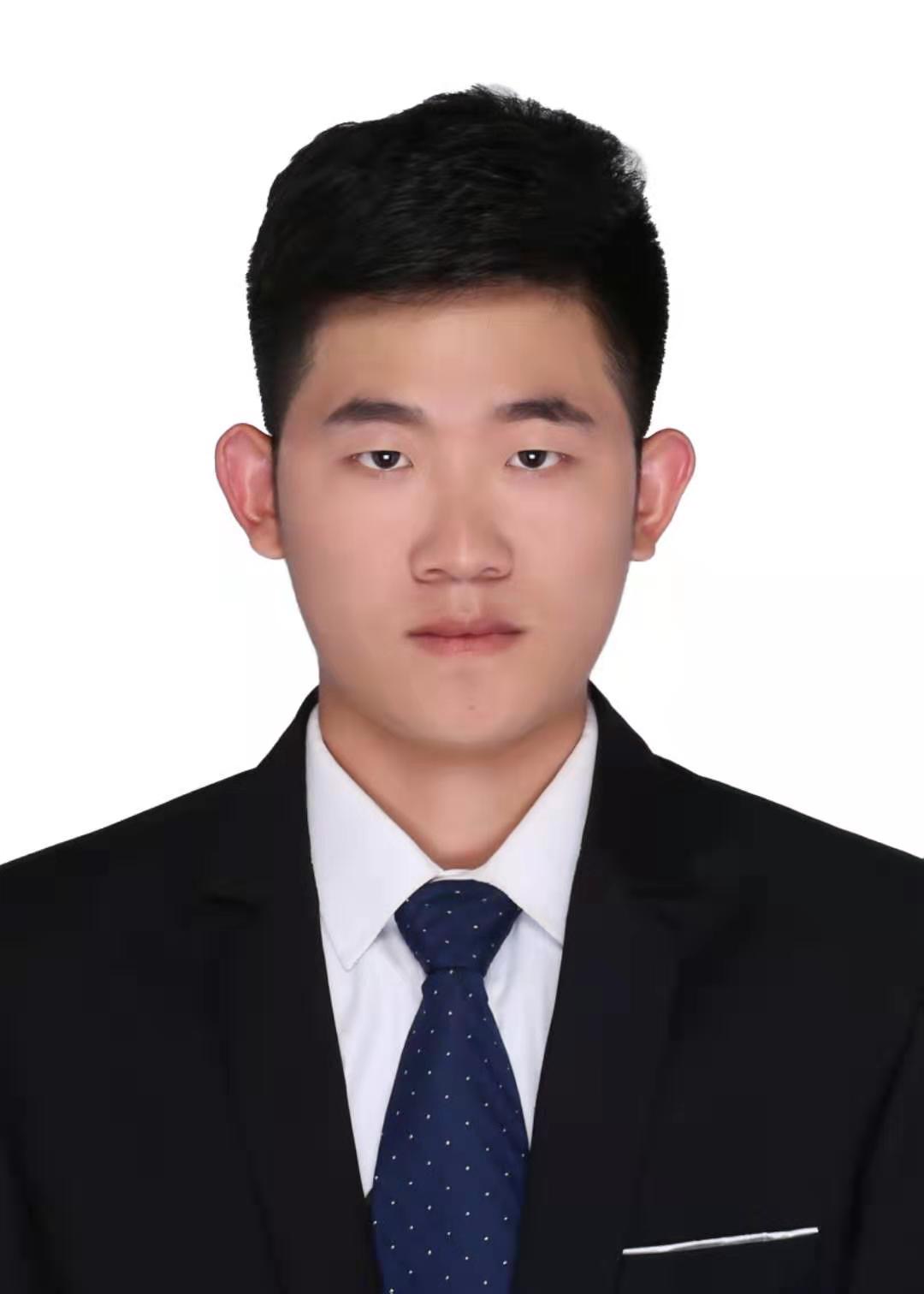}}]{Dongxu Li} received the B.E. degree in communication engineering from University of Electronic Science and Technology of China (UESTC), Chengdu, China, in 2019. He is currently pursuing the Ph.D. degree with the School of Science and Engineering (SSE), The Chinese University of Hong Kong, Shenzhen, China. His current research interests include resonant beam communications and semantic communications.\end{IEEEbiography}

\begin{IEEEbiography}[{\includegraphics[width=1in,height=1.25in,clip,keepaspectratio]{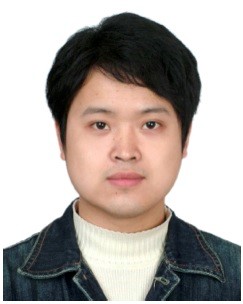}}]{Chuan Huang} (S’09-M’13) received the Ph.D. degree in electrical engineering from Texas A\&M University, College Station, USA, in 2012. From August 2012 to July 2014, he was a Research Associate and then a Research Assistant Professor with Princeton University and Arizona State University, Tempe, respectively. He is currently an Associate Professor with The Chinese University of Hong Kong, Shenzhen. His current research interests include wireless communications and signal processing. He served as a Symposium Chair for IEEE GLOBECOM 2019 and IEEE ICCC 2019 and 2020. He has been serving as an Editor for IEEE TRANSACTIONS ON WIRELESS COMMUNICATIONS, IEEE ACCESS, Journal of Communications and Information Networks, and IEEE WIRELESS COMMUNICATIONS LETTERS.\end{IEEEbiography}

\begin{IEEEbiography}[{\includegraphics[width=1in,height=1.25in,clip,keepaspectratio]{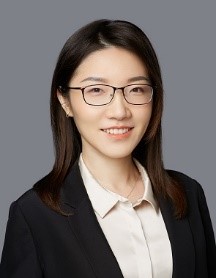}}] {Xiaoqi Qin} (S’13-M’16) received her B.S., M.S., and Ph.D. degrees from Electrical and Computer Engineering with Virginia Tech. She is currently an Associate Professor of School of Information and Communication Engineering with Beijing University of Posts and Telecommunication(BUPT). Her research focuses on exploring performance limits of next-generation wireless networks, and developing innovative solutions for intelligent and efficient machine-type communications. 
\end{IEEEbiography}

\begin{IEEEbiography}[{\includegraphics[width=1in,height=1.25in,clip,keepaspectratio]{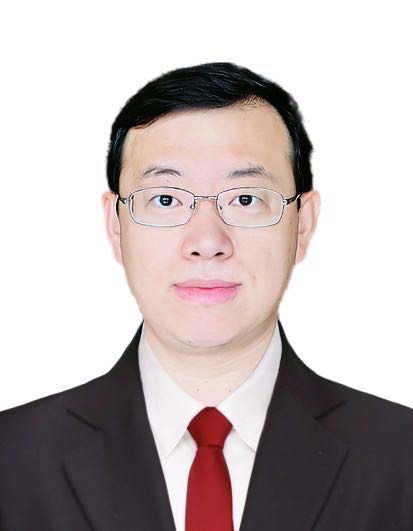}}]{Wei Zhang } (S'01-M'06-SM'11-F'15) received the Ph.D. degree from The Chinese University of Hong Kong in 2005. Currently, he is a Professor at the School of Electrical Engineering and Telecommunications, the University of New South Wales, Sydney, Australia. His current research interests include UAV communications, 5G and beyond. He received 6 best paper awards from IEEE conferences and ComSoc technical committees. He was elevated to Fellow of the IEEE in 2015 and was an IEEE ComSoc Distinguished Lecturer in 2016-2017. 

Within the IEEE ComSoc, he has taken many leadership positions including Member-at-Large on the Board of Governors (2018-2020), Chair of Wireless Communications Technical Committee (2019-2020), Vice Director of Asia Pacific Board (2016-2021), Editor-in-Chief of IEEE Wireless Communications Letters (2016-2019), Technical Program Committee Chair of APCC 2017 and ICCC 2019, Award Committee Chair of Asia Pacific Board and Award Committee Chair of Technical Committee on Cognitive Networks. He was recently elected as Vice President of IEEE Communications Society (2022-2023).

In addition, he has served as a member in various ComSoc boards/standing committees, including Journals Board, Technical Committee Recertification Committee, Finance Standing Committee, Information Technology Committee, Steering Committee of IEEE Transactions on Green Communications and Networking and Steering Committee of IEEE Networking Letters. Currently, he serves as an Area Editor of the IEEE Transactions on Wireless Communications and the Editor-in-Chief of Journal of Communications and Information Networks. Previously, he served as Editor of IEEE Transactions on Communications, IEEE Transactions on Wireless Communications, IEEE Transactions on Cognitive Communications and Networking, and IEEE Journal on Selected Areas in Communications – Cognitive Radio Series.
\end{IEEEbiography}

\end{document}